\documentclass[aps,pra,twocolumn,superscriptaddress,longbibliography,nofootinbib]{revtex4-2}
\usepackage{amsfonts,amsmath,amssymb,epsfig}
\usepackage{epsfig}
\usepackage{graphicx}
\usepackage[dvipsnames]{xcolor}
\usepackage{color}
\usepackage{soul}

\usepackage{epstopdf}
\usepackage{array}
\usepackage{tabularx}
\usepackage[utf8]{inputenc}
\usepackage[T1]{fontenc}
\usepackage[colorlinks,linkcolor=blue, citecolor=blue,
urlcolor=blue]{hyperref}

\begin{document}
	\title{Evaporative damping in open system theory of Bose-Einstein Condensates}
	\author{Nils A. Krause}
	\email{krani857@student.otago.ac.nz}
	\affiliation{Department of Physics, University of Otago, Dunedin, New Zealand}
	\affiliation{Dodd-Walls Centre for Photonic and Quantum Technologies}
	\author{Ashton S. Bradley}
	\affiliation{Department of Physics, University of Otago, Dunedin, New Zealand}
	\affiliation{Dodd-Walls Centre for Photonic and Quantum Technologies}

	\date{\today}
	\begin{abstract}
		We derive a new damping mechanism in the open quantum systems description of Bose-Einstein condensates. It stems from previously neglected terms in the derivation of the stochastic projective Gross-Pitaevskii equation (SPGPE), accounting for a nonlinear evaporation of particles from the coherent into the incoherent region. We demonstrate that the mechanism, while so far assumed to be of minor importance, is comparable in strength to the widely employed number damping. We also provide a simplified (pseudo)-local and a dimensionally reduced form of this evaporative damping. The process completes the SPGPE description of ultracold Bose gases giving a full first-principles picture of their evolution at finite temperature.
	\end{abstract}
	\flushbottom
	\maketitle
	
	\section{Introduction}
	At low temperatures, the dynamics of Bose-Einstein condensates (BECs) can appropriately be described by the Gross-Pitaevskii equation (GPE)\cite{pitaevskii_bose-einstein_2003}. It accurately predicts groundstates\cite{vestergaard_hau_near-resonant_1998}, as well as the existence of non-linear excitations such as solitonic structures\cite{burger_dark_1999,busch_motion_2000,becker_oscillations_2008,jones_motions_1982,meyer_observation_2017} and quantized vortices\cite{madison_vortex_2000,abo-shaeer_observation_2001,kwon_sound_2021}. At higher temperatures closer to the transition point the condensate fraction decreases as particles occupy thermally excited states. Moreover, non-linear excitations are observed to decay\cite{burger_dark_1999,moon_thermal_2015}. The GPE can not account for these processes, nor can it describe condensate formation and growth. Hence, alternative models are required that can capture thermal physics in ultracold Bose gases.
	
\begin{figure}
		\centering
		\includegraphics[width=1\linewidth]{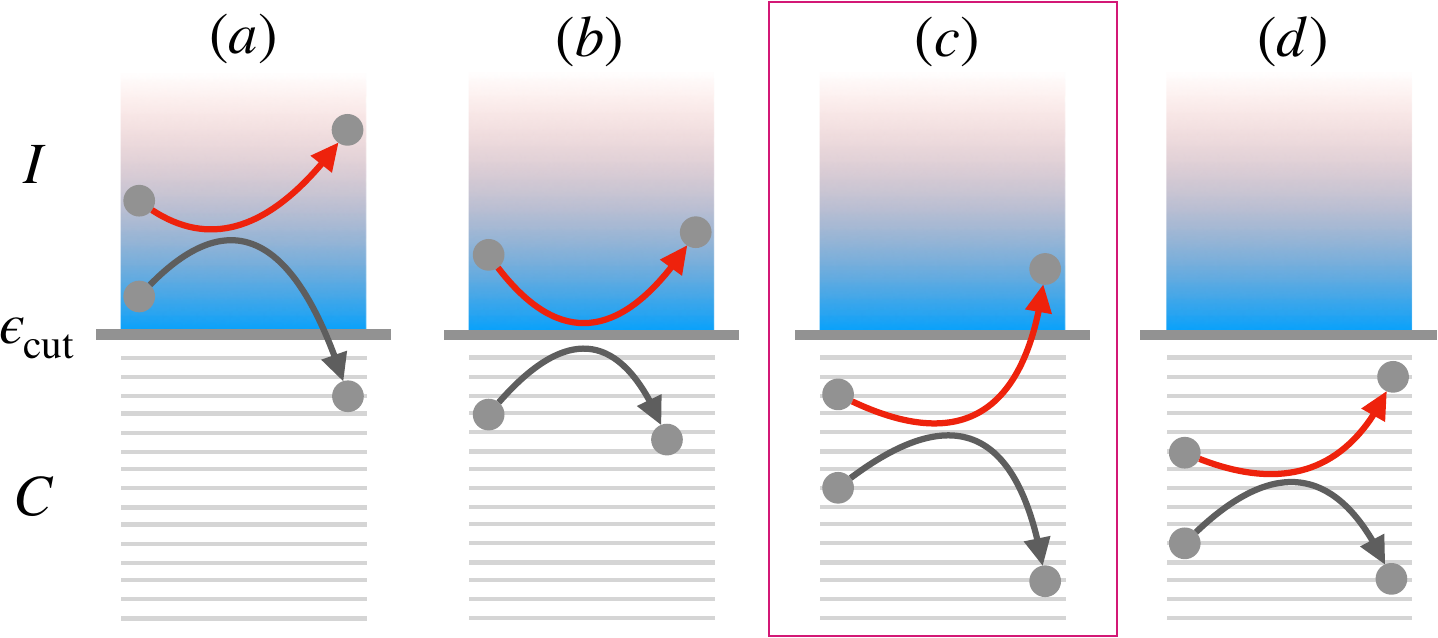}
		\caption{Schematic of collisional damping processes in the SPGPE: (a) Number damping, (b) Energy damping, and (c) Evaporative damping, the focus of the present work. The final process (d) is the Hamiltonian nonlinear interaction in the projected Gross-Pitaevskii evolution describing the coherent dynamics of the classical field. Each process describes an interaction between the coherent region and the thermal cloud involving increasing numbers of $C$-region field operators. The time reverse of each process also occurs.}
		\label{fig:overview}
	\end{figure}

	Several approaches were taken too account for finite temperature effects in BECs \cite{anglin_cold_1997,stoof_initial_1997,stoof_coherent_1999,zaremba_dynamics_1999}. The development of quantum kinetic theory \cite{gardiner_quantum_1997,gardiner_kinetics_1997,jaksch_quantum_1997,gardiner_quantum_1998,jaksch_quantum_1998,gardiner_quantum_2000} and the idea of introducing a projection on a coherent region \cite{davis_dynamics_2001,davis_simulations_2001,davis_simulations_2002} informed the development of the stochastic projected Gross-Pitaevskii equation (SPGPE) \cite{gardiner_stochastic_2002,gardiner_stochastic_2003,blakie_dynamics_2008}. It assumes a classical field (c-field) approximation handling those states with high occupation as a collective classical wave. Particles in high energetic, low occupied states above a cut-off are then treated as a thermal reservoir. Their interaction with the classical wave introduces three processes (schematically depicted in figure \ref{fig:overview}) leading to thermalization of the c-field.
	
	Commonly only one mechanism, so called \emph{number damping} ((a) in figure \ref{fig:overview}), is used to m{o}del dissipation in the SPGPE\cite{bradley_bose-einstein_2008}. It stems from an exchange of particles between the thermal cloud and the coherent region. Number damping has been succesfully applied to calculate and prepare grand-canonical equilibrium distributions \cite{keepfer_phase_2022,underwood_stochastic_2025,rauer_recurrences_2018,ota_collisionless_2018,bayocboc_dynamics_2022}. However, if employed to describe dynamics, it does not match observations and is often arbitrarily enlarged \cite{weiler_spontaneous_2008,kobayashi_thermal_2016,liu_dynamical_2018,liu_kibble-zurek_2020,thudiyangal_universal_2024,comaron_quench_2019,roy_finite-temperature_2021,roy_finite-temperature_2023}.
	
	Recently, the second process in the SPGPE, \emph{energy damping} ((b) in figure \ref{fig:overview}), has attracted more attention. Energy damping describes the scattering between particles in the coherent region and those in the thermal cloud that do not lead to an exchange in particles. Analyzing the process led to the conclusion that at high phase space density energy damping will usually dominate over number damping in both the decay of non-linear structures \cite{mehdi_mutual_2023,krause_thermal_2024} as well as linear relaxation \cite{krause_equilibrium_2025}.
	
	Next to number and energy damping, a third process ((c) in figure \ref{fig:overview}) has been so far discarded in the derivation of the SPGPE as it was claimed to be small\footnote{{A similiar process has been investigated in }\cite{duine_explosion_2001}{ in the context of an explosion of a collapsing BEC. Evaporation out of the condensate was studied, while in the context of the SPGPE we study evaporation out of the coherent region, encompassing more than just the condensate. The process was considered in context of the finite-temperature GPE (FTGPE)}\cite{davis_dynamics_2001}{. However, to our best knwoledge it was never considered in the SPGPE framework.}}. It stems from an evaporation of coherent particles into the incoherent cloud (and its inverse process of a particle capture by the coherent region). {In other words, it describes scattering events between two particles in the coherent region in which one of them gains enough energy to enter the incoherent region, thereby reducing the energy of the other particle.} We derive the effect of this \emph{evaporative damping} on the c-field in the following and conclude that it is not negligible but rather comparable in strength to the number damping process. As most of the physics occurs in the $C$-region, it is driven by highly nonlinear c-field dynamics. This evaporative damping can be cast in a form (\ref{dpsidrift}) that consists of a contribution resembling energy damping (corresponding to the lower particle of figure \ref{fig:overview}(c)) and one that resembles number damping (corresponding to the upper particle of figure \ref{fig:overview}(c)). If we were to choose the cut-off high enough that all trapped particles are in the coherent region, the other mechanisms in the SPGPE, number and energy damping, vanish. Evaporative damping, however, remains, then describing the evaporation of particles out of the trap.
	
	We will begin by reviewing the main assumptions made in the SPGPE and formulating the number and energy damping processes in section \ref{background}. We than derive the evaporative damping in section \ref{3rd}. Its influence on the classical wave is stated in the form of a stochastic equation of motion. Section \ref{Interpretation} gives an interpretation of the different terms appearing in the evaporative damping and estimates its strength with regards to the most commonly employed number damping. In particular, section \ref{Approximate} gives an approximate form of evaporative damping, greatly simplifying its implementation and allowing some analytical progress in its study. Section \ref{DimensionalReduction} derives the equation in a dimensionally reduced form.
	
	\section{Background}
	\label{background}
	Before we calculate the evaporative damping process, we briefly sketch the derivation of the SPGPE and state some of its properties used in the following.
	
	\subsection{SPGPE master equation}
	In the SPGPE \cite{gardiner_stochastic_2003,blakie_dynamics_2008} the field operator $\hat{\Psi}$ describing the entire Bose gas is divided into a high energy part $\hat{\phi}$ and a low energy ($C$-region) part $\hat{\psi}$
	\begin{align}
		\hat{\Psi}(\textbf{r})=\hat{\psi}(\textbf{r})+\hat{\phi}(\textbf{r}),\ \hat{\psi}(\textbf{r})=\sum_n \hat{a}_n\chi_n(\textbf{r}).
	\end{align}
	Here, $\hat{a}_n$ are the annihilation operator of orthonormal states $\chi_n$, which are usually choosen to correspond to those eigenstates of the single particle Hamiltonian
	\begin{align}
		H_\text{sp}=-\frac{\hbar^2}{2m}\Delta +V(\textbf{r})
	\end{align}
	 that have eigenenergy $\epsilon_n$ below a cut-off $\epsilon_\text{cut}$
	 \begin{align}
	 	\label{eigen}
	 	H_\text{sp}\chi_n=\epsilon_n\chi_n,\ \epsilon_n<\epsilon_\text{cut}.
	 \end{align}
	 {In SPGPE treatment, the cut-off should be chosen high enough that eigenstates of the full (interacting) Hamiltonian can be approximated by the eigenstates of the single particle Hamiltonian and low enough that the truncated Wigner approximation is valid. In practice, this leads to a cut-off several times the chemical potential $\mu$}\cite{gardiner_stochastic_2003,krause_equilibrium_2025}. Later, we will also include the influence of the mean field shift due to the interaction energy. The SPGPE is then an evolution equation for a classical wave containing information about the low energy states.

	The evolution of the full system follows the von Neumann equation
	\begin{align}
		\partial_t\rho=-\frac{i}{\hbar}[H,\rho],
	\end{align}
	where the Hamiltonian is given by
	\begin{align}
		\begin{split}
		H&=H_0+H_\text{int},\ H_0=\int d^3\textbf{r}\hat{\Psi}^\dagger(\textbf{r})H_\text{sp}\hat{\Psi}(\textbf{r}),\\
		H_\text{int}&=\frac{g}{2}\int d^3\textbf{r}\hat{\Psi}^\dagger(\textbf{r})\hat{\Psi}^\dagger(\textbf{r})\hat{\Psi}(\textbf{r})\hat{\Psi}(\textbf{r}).
		\end{split}
	\end{align}
	
	Tracing out the high energetic states and thereby introducing the coherent
	\begin{align}
		\rho_C=\text{Tr}_{I}\left\{\rho\right\}
	\end{align}
	and incoherent region
	\begin{align}
		\rho_{I}=\text{Tr}_C\left\{\rho\right\}
	\end{align}
	density operators one arrives under a Markov approximation at an equation with the following contributions: a Hamiltonian term
	\begin{align}
		\partial_t\rho_C|_\text{H}=\frac{i}{\hbar}\hat{L}_C\rho_C,\ \hat{L}_C\rho_C\equiv-[H_C,\rho_C],
	\end{align}
	where $H_C$ contains the forward scattering Hamiltonian $H_\text{forward}$
	\begin{align}
		\begin{split}
		H_C&=\int d^3\textbf{r}\hat{\psi}^\dagger(\textbf{r})H_\text{sp}\hat{\psi}(\textbf{r})\\
		&+\frac{g}{2}\int d^3\textbf{r}\hat{\psi}^\dagger(\textbf{r})\hat{\psi}^\dagger(\textbf{r})\hat{\psi}(\textbf{r})\hat{\psi}(\textbf{r})+H_\text{forward},\\
		H_\text{forward}&=2g\int d^3\textbf{r}n_I(\textbf{r})\hat{\psi}^\dagger(\textbf{r})\hat{\psi}(\textbf{r}).
		\end{split}
	\end{align}
	The subscript $|_H$ refers here and in the following to the evolution induced by the Hamiltonain part and
	\begin{align}
	n_I(\textbf{r})=\text{Tr}_I\{\hat{\phi}^\dagger(\textbf{r})\hat{\phi}(\textbf{r})\rho_I\}
	\end{align}
	is the density of incoherent particles.
	
	Next to the Hamiltonian evolution there are further contributions stemming from the interaction Hamiltonian $H_I$ in the full von Neumann equation that can be divided by the number of $C$-region field operators $\hat{\psi}$, $\hat{\psi}^\dagger$ appearing in $H_I$. Those for which only one appears are called growth terms since they describe particle exchange between thermal cloud and coherent region and lead to the number damping process. Those conatining two are called scattering terms as they describe a particle conserving exchange of energy. They lead to the energy damping process. Lastly, those terms containing three have so far been neglected in the derivation of the SPGPE. Their contribution will be derived in section \ref{3rd}.
	
	The SPGPE assumes that the incoherent (high energetic) region thermalizes fast compared to the coherent region dynamics and the potential is varying slowly in space. Then, it can be taken to be in local equilibrium at all times and thus its Wigner function is given by
	\begin{align}
		\label{F}
		\begin{split}
			F(\textbf{u},\textbf{K})&=\int \frac{d^3\textbf{v}}{(2\pi)^3}e^{i\textbf{K}\cdot\textbf{v}}\langle\hat{\phi}^\dagger(\textbf{u}+\textbf{v}/2)\hat{\phi}(\textbf{u}-\textbf{v}/2)\rangle\\
			&\simeq\frac{\Theta(\hbar\omega(\textbf{u},\textbf{K})-[\epsilon_\text{cut}+2n_\text{tot}(\textbf{u})g])}{e^{\beta[\hbar\omega(\textbf{u},\textbf{K})-\mu]}-1},
		\end{split}
	\end{align}
	where we write here and in the following
	\begin{align}
		\text{Tr}_I\{\cdot\rho_I\}=\langle\cdot\rangle.
	\end{align}
	We also introduced the local energy
	\begin{align}
		\hbar\omega(\textbf{u},\textbf{K})=2n_\text{tot}(\textbf{u})g+V(\textbf{u})+\frac{\hbar^2\textbf{K}^2}{2m}
	\end{align}
	and coherent, incoherent and total density 
	\begin{align}
		\begin{split}
		n_C(\textbf{r})&=\text{Tr}_\text{C}\{\hat{\psi}^\dagger(\textbf{r})\hat{\psi}(\textbf{r})\rho_\text{C}\},\\
		n_I(\textbf{r})&=\text{Tr}_I\{\hat{\phi}^\dagger(\textbf{r})\hat{\phi}(\textbf{r})\rho_I\}=\int_I \frac{d^3\textbf{K}}{(2\pi)^3}F(\textbf{r},\textbf{K}),\\
		n_\text{tot}(\textbf{r})&=n_C(\textbf{r})+n_I(\textbf{r}).
		\end{split}
	\end{align}
	In the integral $I$ stands for integration over the incoherent region, meaning over those momenta $\textbf{K}$ that fullfill
	\begin{align}
		\hbar\omega(\textbf{u},\textbf{K})>\epsilon_\text{cut}+2n_\text{tot}(\textbf{u})g.
	\end{align}
	Contrary to the original derivation \cite{gardiner_stochastic_2003} we include the local (mean-field) interaction energy. Note that under the approximations made in the SPGPE, for the energy damping process all the terms with the interaction energy shift cancel and the results are the same. The inclusion, however, highlights that an explicit neglect is not necessary. For the number damping process we neglect an interaction shift at a later point in the derivation.
	
	In the derivation terms that are time evolved under pure Hamiltonian dynamics appear. In particular, we will need for the incoherent region
	\begin{align}
		\begin{split}
		\langle\hat{\phi}(\textbf{r},0)\hat{\phi}^\dagger(\textbf{r}',\tau)\rangle&\simeq \int_I \frac{d^3\textbf{K}}{(2\pi)^3}[F(\textbf{u},\textbf{K})+1]e^{i\textbf{K}\cdot\textbf{v}+i\omega(\textbf{u},\textbf{K})\tau},\\
		\langle\hat{\phi}^\dagger(\textbf{r},0)\hat{\phi}(\textbf{r}',\tau)\rangle&\simeq\int_I \frac{d^3\textbf{K}}{(2\pi)^3}F(\textbf{u},\textbf{K})e^{-i\textbf{K}\cdot\textbf{v}-i\omega(\textbf{u},\textbf{K})\tau},\\
		\langle\hat{\phi}(\textbf{r},\tau)\hat{\phi}^\dagger(\textbf{r}',0)\rangle&\simeq \int_I \frac{d^3\textbf{K}}{(2\pi)^3}[F(\textbf{u},\textbf{K})+1]e^{i\textbf{K}\cdot\textbf{v}-i\omega(\textbf{u},\textbf{K})\tau},\\
		\langle\hat{\phi}^\dagger(\textbf{r},\tau)\hat{\phi}(\textbf{r}',0)\rangle&\simeq \int_I \frac{d^3\textbf{K}}{(2\pi)^3}F(\textbf{u},\textbf{K})e^{-i\textbf{K}\cdot\textbf{v}+i\omega(\textbf{u},\textbf{K})\tau},
		\end{split}
	\end{align}
	where $\tau$ is a short time and we write $\textbf{u}=(\textbf{r}+\textbf{r}')/2,\ \textbf{v}=\textbf{r}-\textbf{r}'$.	Further, for the coherent region we need
	\begin{align}
		\hat{\psi}(\textbf{r},\tau)=e^{i\hat{L}_C\tau/\hbar}\hat{\psi}(\textbf{r}),\ \hat{\psi}^\dagger(\textbf{r},\tau)=e^{i\hat{L}_C\tau/\hbar}\hat{\psi}^\dagger(\textbf{r}).
	\end{align}
	
	\subsection{Multimode Wigner Representation}
	The next step in the derivation consists in the mapping to a multimode Wigner distribution
	\begin{align}
		\begin{split}
		W_C(\alpha,\alpha^*)&=\int \frac{d^{2M}\lambda}{\pi^{2M}}\exp\left(\sum_m[\lambda_m^*\alpha_m-\lambda_m\alpha_m^*]\right)\\
		&\times\text{Tr}\left\{\rho_C\exp\left(\sum_n[\lambda_n\hat{a}_n^\dagger-\lambda_n^*\hat{a}_n]\right)\right\},
		\end{split}
	\end{align}
	where $M$ is the number of states in the coherent region. Introducing the classical field representation
	\begin{align}
		\psi(\textbf{r})=\sum_n\alpha_n\chi_n(\textbf{r})
	\end{align}
	and the projected functional derivatives
	\begin{align}
		\frac{\delta}{\delta\psi(\textbf{r})}=\sum_n\chi_n^*\frac{\partial}{\partial\alpha_n},\ \frac{\delta}{\delta\psi^*(\textbf{r})}=\sum_n\chi_n\frac{\partial}{\partial\alpha_n^*},
	\end{align}
	terms map in the following way under a multimode Wigner-Weyl transformation:
	\begin{align}
		\begin{split}
		&\hat{\psi}(\textbf{r})\rho_C\leftrightarrow\left(\psi(\textbf{r})+\frac{1}{2}\frac{\delta}{\delta\psi^*(\textbf{r})}\right)W_C,\\
		&\hat{\psi}^\dagger(\textbf{r})\rho_C\leftrightarrow\left(\psi^*(\textbf{r})-\frac{1}{2}\frac{\delta}{\delta\psi(\textbf{r})}\right)W_C \\
		&\rho_C\hat{\psi}(\textbf{r})\leftrightarrow\left(\psi(\textbf{r})-\frac{1}{2}\frac{\delta}{\delta\psi^*(\textbf{r})}\right)W_C,\\
		&\rho_C\hat{\psi}^\dagger(\textbf{r})\leftrightarrow\left(\psi^*(\textbf{r})+\frac{1}{2}\frac{\delta}{\delta\psi(\textbf{r})}\right)W_C.
		\end{split}
	\end{align}
	
	The Hamiltonian part hence maps to
	\begin{align}
		\label{dWH}
		\begin{split}
		&\frac{\partial W_C}{\partial t}\bigg|_H=\int d^3\textbf{r}\left[\frac{i}{\hbar}\frac{\delta}{\delta\psi(\textbf{r})}L_C\psi(\textbf{r})+c.c.\right]W_C\\
		&+\int d^3\textbf{r}\left[\frac{ig}{4\hbar}\frac{\delta^{(2)}}{\delta\psi(\textbf{r})\delta\psi^*(\textbf{r})}\psi^*(\textbf{r})\frac{\delta}{\delta\psi^*(\textbf{r})}+c.c.\right]W_C,
		\end{split}
	\end{align}
	where
	\begin{align}
		L_C=H_\text{sp}+g|\psi(\textbf{r})|^2+2gn_I(\textbf{r}).
	\end{align}
	The second term in (\ref{dWH}) containing a third derivative is dropped under the truncated Wigner approximation (TWA), valid for highly occupied states. Under the same approximation, we will allways only keep the leading order derivative in all the terms appearing in the following. The equation can than be mapped to the stochastic differential equation\footnote{While this part of the equation is deterministic, the mapping from a Wigner function implies a random initial state, hence even under the neglect of the reservoir terms the equation is inherently stochastic.}
	\begin{align}
		i\hbar d\psi(\textbf{r})|_H=\mathcal{P}\{L_C\psi(\textbf{r})dt\}.
	\end{align}
	
	\subsection{Number Damping}
	\label{numbersection}
	Those terms in $H_I$ that contain only one $\hat{\psi}$, $\hat{\psi}^\dagger$ can be seen as describing scattering events under which two particles in the thermal cloud collide with the result of one of the particles losing enough energy to enter the coherent region (and those in which one particle starts of in the coherent region and, via scattering with a particle in the incoherent region, gains enough energy to reach the thermal cloud). These processes are called number damping and noise.
	
	To account fully for the operator $\hat{L}_C$ in this scattering terms is a highly difficult task. Aiming on enabling progres and thereby deriving the SPGPE, an approximation of the form
	\begin{align}
		\label{constLC}
		\begin{split}
		&e^{i(E_1+E_2-E_3-\hat{L}_C)t/\hbar}\hat{\psi}(\textbf{r})\\
		&\sim e^{i(E_1+E_2-E_3-\mu)t/\hbar}\hat{\psi}(\textbf{r})
		\end{split}
	\end{align}
	is performed, where $E_1,E_2,E_3$ are energies of particles in the incoherent region. In other words, the operator $\hat{L}_C$ is approximated by a typical energy scale, namely the cost of adding a particle into the condensate $\mu$\footnote{Strictly, one can be a bit more careful at this point. In practice, one is \emph{less} carefull and approximates further the difference between the interaction energies of the thermal atoms and the chemcial potential by zero, arguing that it should be small compared to the kinetic energies of the thermal particles. In effect, only the damping rate $\gamma$ changes slightly.}.
	
	This leads to an evolution of $\psi$ according to
	\begin{align}
		\label{number}
		i\hbar d\psi(\textbf{r})|_\gamma=\mathcal{P}\left\{-i\gamma[L_C-\mu]\psi(\textbf{r})dt+\sqrt{2\gamma k_\text{B}T\hbar}dW_\gamma\right\}.
	\end{align}
	Here, $dW_\gamma$ is Gaussian complex noise
	\begin{align}
		\langle dW^*_\gamma(\textbf{r},t)dW_\gamma(\textbf{r}',t)\rangle=\delta(\textbf{r}-\textbf{r}')dt
	\end{align}
	and the subscript $|_\gamma$ refers to an evolution according to number damping. Further, we have the dimensionless damping strength\cite{krause_thermal_2024}
	\begin{align}
		\gamma=\frac{8a_\text{s}^2}{\lambda_\text{th}^2}e^{\beta\mu}\int_0^1dy\ln\left(\frac{1-zy}{1-z}\right)\frac{1}{(1-y)(1-zy)},
	\end{align}
	where $z=e^{\beta(\mu-2\epsilon_\text{cut})}$ and we introduced the thermal de Broglie wavelength $\lambda_\text{th}=\sqrt{2\pi\hbar^2/(mk_\text{B}T)}$.
	
	Figure \ref{fig:overview}(a) displays a schematical representation of number damping and noise.
	
	\subsection{Energy Damping}
	Terms in $H_I$ containing two $\hat{\psi}$, $\hat{\psi}^\dagger$ describe the scattering of a particle in the thermal cloud with a particle in the coherent region in which both particles remain in their respective regime but exchange energy and momentum. These mechanisms, dubbed energy damping and noise, induces an evolution according to
	\begin{align}
		\label{energy}
		\begin{split}
		i\hbar(\textbf{S})d\psi(\textbf{r})|_\varepsilon&=\mathcal{P}\bigg\{-\hbar\int d^3\textbf{r}'\varepsilon(\textbf{r}-\textbf{r}')\nabla'\cdot \textbf{j}(\textbf{r}')\psi(\textbf{r})dt\\
		&+\sqrt{2k_\text{B}T\hbar}\psi(\textbf{r})\ dU(\textbf{r})\bigg\}.
		\end{split}
	\end{align}
	The noise terms fullfill the correlations
	\begin{align}
		\langle dU(\textbf{r},t)dU(\textbf{r}',t)\rangle&=\varepsilon(\textbf{r}-\textbf{r}')dt
	\end{align}
	while the subscript $|_\varepsilon$ refers to the evolution induced by energy damping and noise. (\textbf{S}) refers to the equation being in Stratonovich form. The characterising integral kernel is given by\cite{rooney_stochastic_2012}
	\begin{align}
		\label{kernel}
		\varepsilon(\textbf{r})=16\pi a_\text{s}^2N_\text{cut}\int \frac{d^3\textbf{k}}{(2\pi)^3}\frac{e^{i\textbf{k}\cdot\textbf{r}}}{|\textbf{k}|},
	\end{align}
	where $N_\text{cut}=1/[\exp(\beta\{\epsilon_\text{cut}-\mu\})-1]$ is the thermal occupation of states at the cut-off.
	
	Figure \ref{fig:overview}(b) displays a schematic of the processes leading to energy damping and noise.
	
	Next to the processes of number and energy damping a third interaction mechanism between the thermal cloud and the coherent region was previously discarded in the derivation of the SPGPE. We will complete the SPGPE description in the following by deriving the dynamics induced through this third process that we call evaporative damping.
		
	\section{Evaporative Damping}
	\label{3rd}
	In the following we will derive the effect of evaporative damping on the c-field and interpret our findings. We also obtain a simpler approximate (pseudo-)local form.
	
	\subsection{Derivation of the master equation}
	The terms in $H_I$ containing three $\hat{\psi}$, $\hat{\psi}^\dagger$ lead to an evolution of the density matrix given by
	\begin{widetext}
	\begin{align}
		\label{3rdterm}
		\begin{split}
		\partial_t\rho_C|_\sigma=\frac{g^2}{\hbar^2}\int d^3\textbf{r}\int d^3\textbf{r}'\int_{-\infty}^0&d\tau\bigg\{\langle\hat{\phi}(\textbf{r},0)\hat{\phi}^\dagger(\textbf{r}',\tau)\rangle\left[\hat{\psi}^\dagger(\textbf{r}',\tau)\hat{\psi}(\textbf{r}',\tau)\hat{\psi}(\textbf{r}',\tau)\rho_C,\hat{\psi}^\dagger(\textbf{r},0)\hat{\psi}^\dagger(\textbf{r},0)\hat{\psi}(\textbf{r},0)\right]\\
		&+\langle\hat{\phi}^\dagger(\textbf{r}',\tau)\hat{\phi}(\textbf{r},0)\rangle\left[\hat{\psi}^\dagger(\textbf{r},0)\hat{\psi}^\dagger(\textbf{r},0)\hat{\psi}(\textbf{r},0),\rho_C\hat{\psi}^\dagger(\textbf{r}',\tau)\hat{\psi}(\textbf{r}',\tau)\hat{\psi}(\textbf{r}',\tau)\right]\\
		&+\langle\hat{\phi}^\dagger(\textbf{r},0)\hat{\phi}(\textbf{r}',\tau)\rangle\left[\hat{\psi}^\dagger(\textbf{r}',\tau)\hat{\psi}^\dagger(\textbf{r}',\tau)\hat{\psi}(\textbf{r}',\tau)\rho_C,\hat{\psi}^\dagger(\textbf{r},0)\hat{\psi}(\textbf{r},0)\hat{\psi}(\textbf{r},0)\right]\\
		&+\langle\hat{\phi}(\textbf{r}',\tau)\hat{\phi}^\dagger(\textbf{r},0)\rangle\left[\hat{\psi}^\dagger(\textbf{r},0)\hat{\psi}(\textbf{r},0)\hat{\psi}(\textbf{r},0),\rho_C\hat{\psi}^\dagger(\textbf{r}',\tau)\hat{\psi}^\dagger(\textbf{r}',\tau)\hat{\psi}(\textbf{r}',\tau)\right]	\bigg\},
		\end{split}
		\end{align}
		\end{widetext}
		where the subscript $|_\sigma$ refers to the evolution according to evaporative damping. Introducing for readability the third order field operator
		\begin{align}
			\hat{\Phi}(\textbf{r})=\hat{\psi}^\dagger(\textbf{r})\hat{\psi}(\textbf{r})\hat{\psi}(\textbf{r})
		\end{align}
		and the rate functions
		\begin{align}
			\label{Rs}
			\begin{split}
			R^{(-)}\left(\textbf{u},\textbf{v},\hat{L}_C\right)&=\pi\int_I \frac{d^3\textbf{K}}{(2\pi)^3}\left[1+F\left(\textbf{u},\textbf{K}\right)\right]e^{i\textbf{K}\cdot\textbf{v}}\\
			&\times\delta(\omega(\textbf{u},\textbf{K})-\hat{L}_C/\hbar),
		\end{split}
	\end{align}
			\begin{align}
			\begin{split}
			R^{(+)}\left(\textbf{u},\textbf{v},\hat{L}_C\right)&=\pi\int_I \frac{d^3\textbf{K}}{(2\pi)^3}F\left(\textbf{u},\textbf{K}\right)e^{i\textbf{K}\cdot\textbf{v}}\\
			&\times\delta(\omega(\textbf{u},\textbf{K})-\hat{L}_C/\hbar)
			\end{split}
		\end{align}
		we derive\footnote{Approximating $\int_{-\infty}^0 d\tau e^{i\omega\tau}\simeq\pi\delta(\omega)$ (hence neglecting the principal value).}
		\begin{widetext}
		\begin{align}
			\label{master}
			\begin{split}
		\partial_t\rho_C|_\sigma&\simeq\frac{g^2}{\hbar^2}\int d^3\textbf{r}\int d^3\textbf{r}'\bigg\{\left[R^{(-)}\left(\frac{\textbf{r}+\textbf{r}'}{2},\textbf{r}-\textbf{r}',\hat{L}_C\right)\hat{\Phi}(\textbf{r}')\rho_C,\hat{\Phi}^\dagger(\textbf{r})\right]+\left[\hat{\Phi}^\dagger(\textbf{r}'),\rho_CR^{(+)}\left(\frac{\textbf{r}+\textbf{r}'}{2},\textbf{r}-\textbf{r}',\hat{L}_C\right)\hat{\Phi}(\textbf{r})\right]\\
		&+\left[R^{(+)}\left(\frac{\textbf{r}+\textbf{r}'}{2},\textbf{r}-\textbf{r}',-\hat{L}_C\right)\hat{\Phi}^\dagger(\textbf{r}')\rho_C,\hat{\Phi}(\textbf{r})\right]+\left[\hat{\Phi}(\textbf{r}'),\rho_CR^{(-)}\left(\frac{\textbf{r}+\textbf{r}'}{2},\textbf{r}-\textbf{r}',-\hat{L}_C\right)\hat{\Phi}^\dagger(\textbf{r})\right]	\bigg\}.
		\end{split}
	\end{align}
	\end{widetext}
	
	Before we continue the derivation we stop briefly to gain some insight into the master equation (\ref{master}). First, the stationary solution $\rho_\text{s}$ of the master equation is as expected the grand canonical ensemble
	\begin{align}
		\rho_\text{s}\propto\exp(\beta[\mu N_C-H_C]).
	\end{align}
	To see this, we note that due to (\ref{F})
	\begin{align}
		\label{RmRp}
		R^{(-)}\left(\textbf{u},\textbf{v},\hbar\omega\right)=e^{\beta(\hbar\omega-\mu)}R^{(+)}\left(\textbf{u},\textbf{v},\hbar\omega\right).
	\end{align}
	Since
	\begin{align}
		[N_C,\hat{\Phi}]=-\hat{\Phi}
	\end{align}
	we conclude
	\begin{align}
		\begin{split}
		&R^{(-)}\left(\frac{\textbf{r}+\textbf{r}'}{2},\textbf{r}-\textbf{r}',\hat{L}_C\right)\hat{\Phi}(\textbf{r}')\rho_\text{s}\\
		=&\rho_\text{s}R^{(+)}\left(\frac{\textbf{r}+\textbf{r}'}{2},\textbf{r}-\textbf{r}',\hat{L}_C\right)\hat{\Phi}(\textbf{r}').
		\end{split}
	\end{align}
	Replacing in the integrals $\textbf{r}\leftrightarrow\textbf{r}'$ than shows that the grand canonical ensemble is indeed a stationary solution to the master equation.
	
	So far, we did not employ the truncated Wigner approximation. Hence, we can in principal choose a rather high cut-off. We could, for instance, choose all the bound states to be in the coherent region, while the \emph{thermal cloud} contains the free states. Under this choice, the thermal cloud can be assumed to be empty $F(\textbf{u},\textbf{k})=0$ and equation (\ref{master}) describes evaporation of particles out of the trap. Note that then $R^{(+)}$ (which describes particle capture) as well as number and energy damping vanish. The spontaneous evaporative process described by $R^{(-)}$ (see (\ref{Rs})) is the only interaction left with the (empty) bath. Choosing a smaller cut-off the process still describes evaporation, although the evaporated particles now do not leave the trap anymore, instead evaporating into the thermal cloud. Note that the mechanism dubbed evaporative damping in this work also includes a particle capture process described by $R^{(+)}$.
	
	We now continue the derivation of evaporative damping. We simplify the master equation employing the high temperature approximation
	\begin{align}
		\label{exp}
		e^{\beta(\hbar\omega-\mu)}\simeq1+\frac{\hbar\omega-\mu}{k_\text{B}T}.
	\end{align}
	Note that contrary to the number and energy damping situation $\omega$ is not smaller than $\epsilon_\text{cut}/\hbar$, but can rather be up to twice as large. Thus, the truncation of the exponential after the first order introduces a larger error than in the other mechanisms. As long as $\epsilon_\text{cut},\mu\ll k_\text{B}T$ it is nervertheless still a valid approximation.
	
	Under this approximation we find a diffusion term\footnote{We employ the symmetry in the integrand allowing $\textbf{r}\leftrightarrow\textbf{r}'$.}
	\begin{align}
		\label{Diffusion}
		\begin{split}
		&\partial_t\rho_C|_{\sigma,\text{diff}}=-\frac{g^2}{\hbar^2}\int d^3\textbf{r}\int d^3\textbf{r}'\\
		&\left\{\left[\hat{\Phi}^\dagger(\textbf{r}),\left[R^{(+)}\left(\frac{\textbf{r}+\textbf{r}'}{2},\textbf{r}-\textbf{r}',\hat{L}_C\right)\hat{\Phi}(\textbf{r}'),\rho_C\right]\right]+h.c.\right\}
		\end{split}
	\end{align}
	and a damping term
	\begin{align}
		\label{Drift}
		\begin{split}
		&\partial_t\rho_C|_{\sigma,\text{drift}}=-\frac{g^2}{\hbar^2 k_\text{B}T}\int d^3\textbf{r}\int d^3\textbf{r}'\bigg\{\\
		&\left[(\mu-\hat{L}_C)R^{(+)}\left(\frac{\textbf{r}+\textbf{r}'}{2},\textbf{r}-\textbf{r}',\hat{L}_C\right)\hat{\Phi}(\textbf{r}')\rho_C,\hat{\Phi}^\dagger(\textbf{r})\right]\\
		&+h.c.\bigg\}.
		\end{split}
	\end{align}
	
	\subsection{Energy scale}
	Similiar to the number damping case \ref{numbersection} it is a highly difficult task to account properly for the operator $\hat{L}_C$ in $R^{(+)}$. Under the analogous approximation as for number damping (\ref{constLC}) both the diffusion and the damping vanish, which can lead to the fallacy that evaporative damping would not be a relevant contribution. However, in contrast to number and energy damping, now the energy required for moving a particle from the coherent into the incoherent region (or vice versa) can not be compensated by an energy shift of a particle staying in the incoherent region and has to be compensated by the energy stored in the coherent region. Hence, we typically have $|\omega|>\omega_\text{cut}$, where $\hbar\omega_\text{cut}=\epsilon_\text{cut}+2n_\text{tot}g$. Indeed, an argumentation similiar to the one for number damping presented in appendix \ref{energycon} suggests that the relevant energies lie in the range $\epsilon_\text{cut}+2n_\text{tot}g$ to $2\epsilon_\text{cut}+2n_\text{tot}g$.
	Therefore, the energy exchange in the coherent region is on the scale of the energies of particles in the incoherent region and it is far from clear that neglecting this energy exchange is still a valid simplification. To enable progress, we proceed along similiar lines to the number damping case, replacing the operator by a typical energy scale. This time, however, the natural choice is the smallest possible energy exchange
	\begin{align}
		\label{Rp}
		\begin{split}
		R^{(+)}(\textbf{u},\textbf{v},\hbar\omega)&\simeq R^{(+)}(\textbf{u},\textbf{v},\hbar\omega_\text{cut})\\
		&=\frac{1}{2\pi}\frac{mk_\text{cut}(\textbf{u})}{\hbar}N_\text{cut}\text{sinc}(k_\text{cut}(\textbf{u})|\textbf{v}|),
		\end{split}
	\end{align}
	where sinc denotes the unnormalized sinc function
	\begin{align}
		\text{sinc}(x)=\frac{\sin(x)}{x}.
	\end{align}
	This is, we assume that the {majority of} particle{s} in the incoherent region participating in the scattering ha{ve} an energy just above the cut-off, where the thermal distribution is highest. The particles in the coherent region likewise are then choosen from those states that are highest populated, while still carrying enough energy for the evaporation. It also implies that we weaken the strict energy conservation in the scattering, the implicit momentum conservation, however, remains. {In the following we argue that} setting $\omega$ to $\omega_\text{cut}$ in $R^{(+)}$ is justified{ while we need to consider the} exponential up to linear order (\ref{exp}). To see this we expand $R^{(+)}$ in lowest order around $\omega_\text{cut}$ (for $\hbar\omega\geq\hbar\omega_\text{cut}$)
	\begin{align}
		\begin{split}
			&R^{(+)}(\textbf{u},\textbf{v},\hbar\omega)\\
			&\simeq R^{(+)}(\textbf{u},\textbf{v},\hbar\omega_\text{cut})+[\omega-\omega_\text{cut}]\partial_\omega R^{(+)}(\textbf{u},\textbf{v},\hbar\omega)|_{\omega=\omega_\text{cut}},
		\end{split}
	\end{align}
	where
	\begin{align}
		\begin{split}
		\partial_\omega &R^{(+)}(\textbf{u},\textbf{v},\hbar\omega)|_{\omega=\omega_\text{cut}}\\&=[N_\text{cut}R^{(-)}(\textbf{u},\textbf{v},\hbar\omega_\text{cut})\beta\hbar\\
		&+\text{cot}(k_\text{cut}|\textbf{v}|)k_\text{cut}|\textbf{v}|\frac{m}{\hbar k_\text{cut}^2}R^{(+)}(\textbf{u},\textbf{v},\hbar\omega_\text{cut})].
		\end{split}
	\end{align}
	Expanding $R^{(-)}$ around $\omega_\text{cut}$ we find using (\ref{RmRp})
	\begin{align}
		\begin{split}
			&R^{(-)}(\textbf{u},\textbf{v},\hbar\omega)\simeq [1+\beta(\hbar\omega-\mu)]R^{(+)}\left(\textbf{u},\textbf{v},\hbar\omega\right)\\
			&\simeq R^{(+)}\left(\textbf{u},\textbf{v},\hbar\omega_\text{cut}\right)+\beta[\hbar\omega-\mu]R^{(+)}\left(\textbf{u},\textbf{v},\hbar\omega_\text{cut}\right)\\
			&+[\omega-\omega_\text{cut}]\partial_\omega R^{(+)}(\textbf{u},\textbf{v},\hbar\omega)|_{\omega=\omega_\text{cut}}.
		\end{split}
	\end{align}
	Since we expect {the energies of the evaporated particles to be close to the cut-off we have} $\omega-\omega_\text{cut}\ll\omega-\mu/\hbar$ {and} we conclude that the approximations
	\begin{align}
		\label{Rapprox}
		\begin{split}
			&R^{(+)}(\textbf{u},\textbf{v},\hbar\omega)\simeq R^{(+)}(\textbf{u},\textbf{v},\hbar\omega_\text{cut}),\\ &R^{(-)}(\textbf{u},\textbf{v},\hbar\omega)\simeq R^{(+)}(\textbf{u},\textbf{v},\hbar\omega_\text{cut})[1+\beta\hbar\{\omega-\mu/\hbar\}]
		\end{split}
	\end{align}
	give the leading order in $\omega-\omega_\text{cut}$ and $\beta\omega$.
	
	We calculate the resulting diffusion and damping under the approximations (\ref{Rapprox}) in the following.
	
	\subsection{Phase space picture}
	Inserting the constant energy approximation into (\ref{Diffusion}) and (\ref{Drift}) allows us to proceed in the derivation. We first consider the diffusion (\ref{Diffusion}). Performing a Wigner-Weyl transformation and keeping only terms with second order derivatives\footnote{0th and 1st order derivatives vanish. Higher order derivatives are neglected in consistency with the TWA.} we have
	\begin{align}
		\label{dWdiff}
		\begin{split}
		&\partial_tW_C|_{\sigma,\text{diff}}=\frac{1}{2\pi}\frac{mg^2}{\hbar^3}\int d^3\textbf{r}\int d^3\textbf{r}'k_\text{cut}(\textbf{u})N_\text{cut}\\
		&\times\text{sinc}\left(k_\text{cut}|\textbf{r}-\textbf{r}'|\right)\bigg\{\frac{\delta}{\delta\psi(\textbf{r})}\psi(\textbf{r})\psi(\textbf{r})\frac{\delta}{\delta\psi^*(\textbf{r}')}\psi^*(\textbf{r}')\psi^*(\textbf{r}')\\
		&-2\frac{\delta}{\delta\psi(\textbf{r})}\psi(\textbf{r})\psi(\textbf{r})\frac{\delta}{\delta\psi(\textbf{r}')}\psi^*(\textbf{r}')\psi(\textbf{r}')\\
		&-2\frac{\delta}{\delta\psi(\textbf{r})}\psi^*(\textbf{r})\psi(\textbf{r})\frac{\delta}{\delta\psi(\textbf{r}')}\psi(\textbf{r}')\psi(\textbf{r}')\\
		&+4\frac{\delta}{\delta\psi(\textbf{r})}\psi^*(\textbf{r})\psi(\textbf{r})\frac{\delta}{\delta\psi^*(\textbf{r}')}\psi^*(\textbf{r}')\psi(\textbf{r}')+h.c.\bigg\}W_C.
		\end{split}
	\end{align}
	This maps to the SPGPE contribution (see appendix \ref{mapping})
	\begin{align}
		\label{dpsidiff}
		(\textbf{S})i\hbar d\psi|_{\sigma,\text{diff}}=\mathcal{P}\{i\hbar\psi^2 dW-2i\hbar|\psi|^2 dW^*\},
	\end{align}
	where
	\begin{align}
		\langle dW(\textbf{r})dW^*(\textbf{r}')\rangle=\frac{mg^2k_\text{cut}(\textbf{u})}{\pi\hbar^3}N_\text{cut}\text{sinc}\left(k_\text{cut}|\textbf{r}-\textbf{r}'|\right)dt.
	\end{align}
	
	We now consider the damping (\ref{Drift}). We obtain by only keeping first order derivatives
	\begin{align}
		\begin{split}
		&\partial_tW_C|_{\sigma,\text{drift}}=-\frac{1}{2\pi}\frac{mg^2}{\hbar^3k_\text{B}T}\int d^3\textbf{r}\int d^3\textbf{r}'k_\text{cut}(\textbf{u})N_\text{cut}\\
		&\times\text{sinc}\left(k_\text{cut}|\textbf{r}-\textbf{r}'|\right)\bigg\{\bigg[2\frac{\delta}{\delta\psi(\textbf{r})}\psi^*(\textbf{r})\psi(\textbf{r})\\
		&-\frac{\delta}{\delta\psi^*(\textbf{r})}\psi^*(\textbf{r})\psi^*(\textbf{r})\bigg][\mu\psi^*(\textbf{r}')\psi(\textbf{r}')\psi(\textbf{r}')-g|\psi(\textbf{r}')|^4\psi(\textbf{r}')\\
		&+\psi(\textbf{r}')^2H_\text{sp}\psi^*(\textbf{r}')-2|\psi(\textbf{r}')|^2H_\text{sp}\psi(\textbf{r}')]+h.c.\bigg\}W_C.
		\end{split}
	\end{align}
	This maps to the SPGPE contribution
	\begin{align}
		\begin{split}
		&d\psi(\textbf{r})|_{\sigma,\text{drift}}=\mathcal{P}\bigg\{\frac{1}{2\pi}\frac{mg^2}{\hbar^3k_\text{B}T}\int d^3\textbf{r}'k_\text{cut}(\textbf{u})N_\text{cut}\\
		&\times\text{sinc}\left(k_\text{cut}|\textbf{r}-\textbf{r}'|\right)\big\{2|\psi(\textbf{r})|^2\times\big[\mu|\psi(\textbf{r}')|^2\psi(\textbf{r}')\\
		&-g|\psi(\textbf{r}')|^4\psi(\textbf{r}')+\psi(\textbf{r}')^2H_\text{sp}\psi^*(\textbf{r}')-2|\psi(\textbf{r}')|^2H_\text{sp}\psi(\textbf{r}')\big]\\
		&-(\psi(\textbf{r}))^2\times\big[\mu|\psi(\textbf{r}')|^2\psi^*(\textbf{r}')-g|\psi(\textbf{r}')|^4\psi^*(\textbf{r}')\\
		&+\psi^*(\textbf{r}')^2H_\text{sp}\psi(\textbf{r}')-2|\psi(\textbf{r}')|^2H_\text{sp}\psi^*(\textbf{r}')\big]\big\}\bigg\}dt.
		\end{split}
	\end{align}
	
	In order to write the diffusion and drift contributions together more compactly we introduce the dimensionless function characterising the strength of the process
	\begin{align}
		\begin{split}
			&\sigma(\textbf{r},\textbf{r}')\\
			&=\frac{mg^2N_\text{cut}}{2\pi\hbar^2k_\text{B}T}k_\text{cut}(\textbf{u})\text{sinc}\left(k_\text{cut}\left(\textbf{u}\right)|\textbf{r}-\textbf{r}'|\right)\psi^*(\textbf{r})\psi(\textbf{r}')\\
			&=4a_\text{s}^2\lambda_\text{th}^2N_\text{cut}k_\text{cut}(\textbf{u})\text{sinc}\left(k_\text{cut}\left(\textbf{u}\right)|\textbf{r}-\textbf{r}'|\right)\psi^*(\textbf{r})\psi(\textbf{r}').
		\end{split}
	\end{align}
With this definition, we find
	\begin{widetext}
	\begin{align}
		\label{dpsidrift}
		\begin{split}
			i\hbar (\textbf{S})d\psi(\textbf{r})|_{\sigma}&=\mathcal{P}\bigg\{-i\int d^3\textbf{r}'\text{Re}\{\sigma(\textbf{r},\textbf{r}')\}\psi(\textbf{r})\psi^*(\textbf{r}')[L_C-\mu]\psi(\textbf{r}')+3\int d^3\textbf{r}'\text{Im}\{\sigma(\textbf{r},\textbf{r}')\}\psi(\textbf{r})\psi^*(\textbf{r}')[L_C-\mu]\psi(\textbf{r}')\\
			&-4\hbar\psi(\textbf{r})\int d^3\textbf{r}'\text{Re}\{\sigma(\textbf{r},\textbf{r}')\}\nabla'\cdot\textbf{j}(\textbf{r}')\bigg\}dt+\mathcal{P}\left\{i\hbar\psi dw+\hbar\psi dU\right\},
		\end{split}
	\end{align}
	where we introduced the noises
	\begin{align}
		dw\equiv-\psi^* dW^*,\ dU\equiv2\text{Im}\{i\psi dW\}=2\text{Im}\{idw\},
	\end{align}
	fullfilling the correlations
	\begin{align}
			\langle dw(\textbf{r})dw^*(\textbf{r}')\rangle=\frac{2k_\text{B}T}{\hbar}\sigma(\textbf{r},\textbf{r}')dt,\ \langle dU(\textbf{r})dU(\textbf{r}')\rangle=\frac{4k_\text{B}T}{\hbar}\sigma(\textbf{r},\textbf{r}')dt,\ \langle dU(\textbf{r})dw^*(\textbf{r}')\rangle=-\frac{2k_\text{B}T}{\hbar}\sigma(\textbf{r},\textbf{r}')dt.
	\end{align}
	\end{widetext}
	
	With equation (\ref{dpsidrift}) we derived an equation for the evaporative damping describing terms in the interaction Hamiltonian $H_I$ containing three c-field operators $\hat{\psi}$, $\hat{\psi}^\dagger$. Evaporative damping can be decomposed into two processes remeniscent of number and energy damping as discussed in the next section.
	
	\section{Importance of evaporative damping}
	\label{Interpretation}
	In the previous section we derived a stochastic differential equation capturing the effect of evaporative damping on the c-field. This section explores the meaning of the terms appearing and gives a simplified version of evaporative damping that allows for an estimate of its importance in the dynamics of a Bose gas.
	
	\subsection{Interpretation}
	Equation (\ref{dpsidrift}) describes the full evaporative damping and noise. Several terms appear that, considered isolatedly, behave similiarly to the known processes of number and energy damping: the first term
	\begin{align}
		\label{numberlike}
		\begin{split}
			i\hbar (\textbf{S})d\psi(\textbf{r})|_{\sigma,1}&=\mathcal{P}\bigg\{-i\int d^3\textbf{r}'\text{Re}\{\sigma(\textbf{r},\textbf{r}')\}\psi(\textbf{r})\psi^*(\textbf{r}')\\
			&\times[L_C-\mu]\psi(\textbf{r}')dt+i\hbar\psi dw\bigg\}
			\end{split}
	\end{align}
	corresponds to a number damping (c.f. equation (\ref{number})) like contribution. It describes particles that enter the coherent region from the incoherent region (and vice versa) due to a scattering event with a particle staying in the coherent region. Consequently, it is this term that is responsible for particle fluctuations in the $C$-region caused by evaporative damping, see appendix \ref{numbergrowth}. The terms
	\begin{align}
		\label{energylike}
		\begin{split}
			&i\hbar (\textbf{S})d\psi(\textbf{r})|_{\sigma,2}\\
			&=\mathcal{P}\left\{-4\hbar\psi(\textbf{r})\int d^3\textbf{r}'\text{Re}\{\sigma(\textbf{r},\textbf{r}')\}\nabla'\cdot\textbf{j}(\textbf{r}')dt+\hbar\psi dU\right\}
			\end{split}
	\end{align}
	correspond to an energy damping (c.f. equation (\ref{energy})) like contribution stemming from the damping of the particles staying in the coherent region under the scattering event. This expression is number conserving and only describes energy transfer between the coherent region and the thermal cloud. The last term
	\begin{align}
		\begin{split}
			&i\hbar d\psi(\textbf{r})|_{\sigma,3}\\
			&=\mathcal{P}\left\{3\int d^3\textbf{r}'\text{Im}\{\sigma(\textbf{r},\textbf{r}')\}\psi(\textbf{r})\psi^*(\textbf{r}')[L_C-\mu]\psi(\textbf{r}')dt\right\}
			\end{split}
	\end{align}
	as well as the correlations between the two noise terms than take into account the interdependence between these events. This interpretation becomes more apparent in an approximate form that we derive in the next section.
	
	Figure \ref{fig:overview}(c) shows a schematic of the processes that leads to the identified evolution. The number damping like term identified in (\ref{dpsidrift}) (equation (\ref{numberlike})) corresponds to the upper part of the scattering events (exchange of particle between coherent and incoherent region) considered isolatedly, while the energy damping like term (\ref{energylike}) corresponds to the lower part considered isolatedly. In the following, we analyze the importance of evaporative damping approximating the found dynamics.
	
	\subsection{Approximate (Pseudo-)local form}
	\label{Approximate}
	For temperatures above the critical temperature there is only short length coherence. Hence, $\psi^*(\textbf{r})\psi(\textbf{r}')$ quickly approaches zero with increasing $\text{{\textbf{v}=}}\textbf{r}-\textbf{r}'$ except for random fluctuations and we can assume that $\sigma$ is sharp in $\text{{\textbf{v}}}$. Thus, in the other terms in the integrals $\textbf{r}'\simeq\textbf{r}$ and $\text{Im}\{\sigma\}\simeq0$. We can argue similarly below the critical temperature if the trap is approximately flat on the scale of the healing length $\xi=\hbar/\sqrt{mn_Cg}$. Then, for $|\text{{\textbf{v}}}|\gg\xi$ but $|\text{{\textbf{v}}}|$ small compared to the trap extension, $\langle\psi^*(\textbf{r})\psi(\text{{$\textbf{r}'$}})\rangle$ is a constant and considering that the integral over the sinc gives zero we can replace $\psi^*(\textbf{r})\psi(\textbf{r}')\rightarrow\psi^*(\textbf{r})\psi(\textbf{r}')-\langle\psi^*(\textbf{r})\psi(\text{{$\textbf{r}'$}})\rangle$. We are then again left with only fluctuations for $|\textbf{r}-\textbf{r}'|\gg\xi$. Thus, the approximations are possible at all temperatures. We define
	\begin{align}
		\bar{\sigma}(\textbf{r})=n_p\int d^3\textbf{r}'\text{Re}\{\sigma(\textbf{r},\textbf{r}')\},
	\end{align}
	where $n_p$ denotes the peak density
	\begin{align}
		n_p=\max_\textbf{r}\langle|\psi(\textbf{r})|^2\rangle.
	\end{align}
	Then, the equation takes the (pseudo-)local form
	\begin{widetext}
	\begin{align}
		\label{pseudo-local}
		\begin{split}
			i\hbar (\textbf{S})d\psi(\textbf{r})|_\sigma&=\mathcal{P}\left\{-i\bar{\sigma}(\textbf{r})\frac{|\psi(\textbf{r})|^2}{n_p}[L_C-\mu]\psi(\textbf{r})dt+i\hbar\psi dw-4\frac{\bar{\sigma}(\textbf{r})}{n_p}\hbar[\nabla\cdot\textbf{j}(\textbf{r})]\psi(\textbf{r})dt+\hbar\psi dU\right\},
		\end{split}
	\end{align}
	where the noises fullfill the correlations
	\begin{align}
		\begin{split}
		\langle dw(\textbf{r})dw^*(\textbf{r}')\rangle&=\frac{2k_\text{B}T}{\hbar}\frac{\bar{\sigma}(\textbf{r})}{n_p}\delta(\textbf{r}-\textbf{r}')dt,\ \langle dU(\textbf{r})dU(\textbf{r}')\rangle=\frac{2k_\text{B}T}{\hbar}\frac{2\bar{\sigma}(\textbf{r})}{n_p}\delta(\textbf{r}-\textbf{r}')dt,\\
		\langle dU(\textbf{r})dw^*(\textbf{r}')\rangle&=-\frac{2k_\text{B}T}{\hbar}\frac{\bar{\sigma}(\textbf{r})}{n_p}\delta(\textbf{r}-\textbf{r}')dt.
		\end{split}
	\end{align}
	\end{widetext}
	Note that the drift term describing the interdependence between the processes identified as number and energy damping like vanishes. Their relation manifests now solely in the correlation of the noise terms.
	
	 Assuming that the fluctuations in the integral over $\sigma$ cancel we can employ the (in general very rough) approximation
	\begin{align}
		\psi(\textbf{r})\psi^*(\textbf{r}')\sim \langle\psi(\textbf{r})\psi^*(\textbf{r}')\rangle.
	\end{align}
	In a box $k_\text{cut}(\textbf{r})=k_\text{cut}$ and in equilibrium the average depends only on $v=|\textbf{r}-\textbf{r}'|$ and we have $n_p=n_C$. We conclude
	\begin{align}
		\label{sigmabar}
		\bar{\sigma}(\textbf{r})=\frac{N_\text{cut}}{\pi}\frac{\lambda_\text{th}^2}{\xi^2}\int_0^\infty \frac{dv}{\xi}\frac{v}{\xi}\sin(k_\text{cut}v)\frac{\langle\psi(\textbf{r})\psi^*(\textbf{r}')\rangle}{n_C}.
	\end{align}
	This (now truely local) form, while derived under rather crude approximations, allows a straightforward estimation of the strength of evaporative damping and comparison to number damping. Note also that under the further approximation $|\psi|^2\sim n_C$ the number damping like term (\ref{numberlike}) takes \emph{the same} form as number damping, with only the pre-factor differing. We can estimate $\bar{\sigma}$ in three different temperature regimes.
	
	 Far below the critical temperature (at high phase space density $n_C\lambda_\text{th}^3\gg1$) phase fluctuations dominate over density fluctuations. As we are only looking for a rough estimate of $\bar{\sigma}$ we neglect the latter. Phase fluctuations are characterised by the variance\cite{krause_equilibrium_2025}
	\begin{align}
		\label{thetavar}
		-\frac{\text{Var}[\theta(\textbf{v})]}{2}=\int \frac{d^3\textbf{k}'}{(2\pi)^3}\frac{2\pi}{n_C\lambda_\text{th}^2\xi}\frac{e^{i\textbf{k}'\cdot\textbf{v}}-1}{(k')^2}\Theta(k_\text{cut}-k')
	\end{align}
	yielding
	\begin{align}
		\bar{\sigma}(\textbf{r})=\frac{1}{48}\frac{N_\text{cut}}{n_C\lambda_\text{th}^2\xi n_C\xi^3}=\frac{\pi^2}{3}N_\text{cut}\frac{a_\text{s}^2}{\lambda_\text{th}^2}.
	\end{align}
	\begin{figure}
		\centering
		\includegraphics[width=1\linewidth]{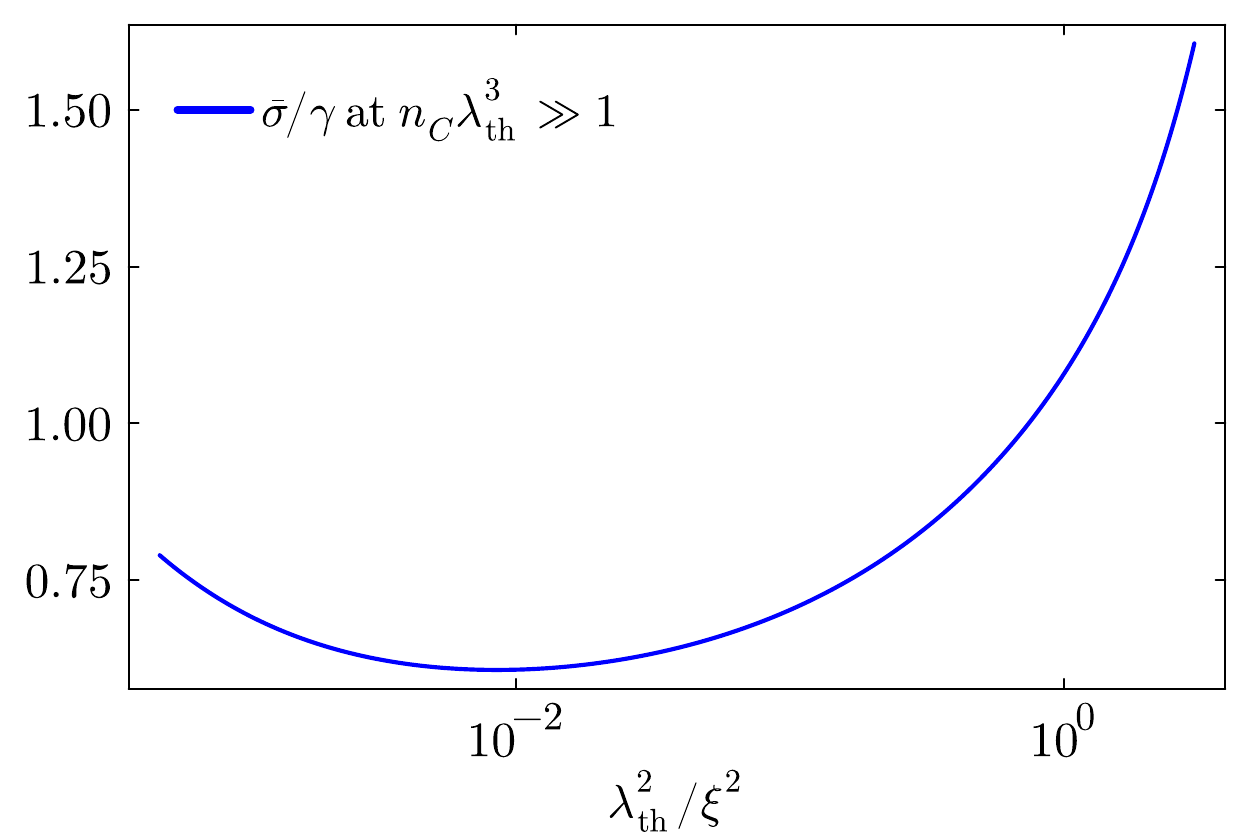}
		\caption{Number damping strength $\gamma$ vs evaporative damping strength $\bar{\sigma}$ in the high phase space density regime. Number damping and evaporative damping are of comparable size.}
		\label{fig:sigmavsgam}
	\end{figure}
	Thus, at high phase space density evaporative damping will usually be comparable to number damping which is determined by the parameter (for $N_\text{cut}\geq2$)\cite{krause_equilibrium_2025}
	\begin{align}
		\label{gambound}
		\gamma<\frac{4\pi^2}{3}N_\text{cut}\frac{a_\text{s}^2}{\lambda_\text{th}^2}.
	\end{align}
	See also figure \ref{fig:sigmavsgam} for a comparison of $\gamma$ and $\bar{\sigma}$ in this regime. A more thoroughly study of evaporative damping at high phase-space density is provided in appendix \ref{Linearization}.
	
	Close to the critical temperature (and hence at phase space density of order one $n_C\lambda_\text{th}^3\sim1$\footnote{{This assumption requires that the coherent density is still comparable to the total density, which in the SPGPE close to the transition point is typically true.}}) the coherence length will be on the order of a healing length and the integral hence on the order of one, so that we have in this regime
	\begin{align}
		\label{sigmanl1}
		\bar{\sigma}(\textbf{r})\sim\frac{N_\text{cut}}{\pi}\frac{\lambda_\text{th}^2}{\xi^2}=4N_\text{cut}n_C\lambda_\text{th}^3\frac{a_\text{s}}{\lambda_\text{th}}\sim4N_\text{cut}\frac{a_\text{s}}{\lambda_\text{th}}.
	\end{align}
	As long as the interactions are weak enough so that $a_\text{s}\ll\lambda_\text{th}$, evaporative damping will thus typically dominate over number damping close to the critical temperature.
	
	Above the critical temperature ($n_C\lambda_\text{th}^3\ll1$), we expect not much coherence between states and can hence write with the average occupation of states $N_\textbf{k}=\langle\tilde{\psi}(\textbf{k})\tilde{\psi}^*(\textbf{k}')\rangle d^3\textbf{k}'$
	\begin{align}
		\begin{split}
		\langle\psi(\textbf{r})\psi^*(\textbf{r}')\rangle&=\int d^3\textbf{k}\int d^3\textbf{k}'e^{i\textbf{k}\cdot\textbf{r}}e^{-i\textbf{k}'\cdot\textbf{r}'}\langle\tilde{\psi}(\textbf{k})\tilde{\psi}^*(\textbf{k}')\rangle\\
		&\simeq\int d^3\textbf{k}N_\textbf{k}e^{i\textbf{k}\cdot(\textbf{r}-\textbf{r}')}\\
		&=4\pi\int_0^\infty dkk^2\text{sinc}(kv)N_k.
		\end{split}
	\end{align}
	The integral in (\ref{sigmabar}) can thus be written as
	\begin{align}
		\begin{split}
		&\int_0^\infty \frac{dv}{\xi}\frac{v}{\xi}\sin(k_\text{cut}v)\frac{\langle\psi(\textbf{r})\psi^*(\textbf{r}')\rangle}{n_C}\\
		&\simeq \frac{2\pi^2 k_\text{cut}}{n_C\xi^2}N_\text{cut}
		\end{split}
	\end{align}
	so that we conclude
	\begin{align}
		\bar{\sigma}\sim2\pi N_\text{cut}^2\frac{k_\text{cut}\lambda_\text{th}^2}{n_C\xi^4}=\frac{(2\pi)^3}{2}n_C\lambda_\text{th}^3k_\text{cut}\lambda_\text{th}8N_\text{cut}^2\frac{a_\text{s}^2}{\lambda_\text{th}^2}.
	\end{align}
	Note that \cite{krause_thermal_2024}
	\begin{align}
		\gamma<8N_\text{cut}^2\frac{a_\text{s}^2}{\lambda_\text{th}^2}.
	\end{align}
	Thus, as long as $n_C\lambda_\text{th}^3\gtrsim1/100$ we can still expect
	\begin{align}
		\gamma<\bar{\sigma}
	\end{align}
	and only for temperatures far above the critical temperature number damping dominates.

\section{Dimensional Reduction}
\label{DimensionalReduction}
We derived the evaporative damping mechanism in three dimensions and analyzed its importance. In the following we will argue analogously to the three dimensional case to derive a two- and one-dimensional reduced form of the evaporative damping. We start with equation (\ref{3rdterm}). We consider the situation in which along one or two axes a deep harmonic trap is applied. We assume that all states in the coherent region are along the axes we reduce in the single particle ground state.{ Note that, in contrast to the dimensional reduction performed for number and energy damping}\cite{bradley_low-dimensional_2015}{, we do not require the thermal cloud to be three dimensional. The results in this section hence extend into the regime of deep traps. }We can thus write in the two and one dimensional case
\begin{align}
	\hat{\psi}(\textbf{r})=\hat{\psi}(\textbf{r}_\perp)\varphi_z(z),\ \hat{\psi}(\textbf{r})=\hat{\psi}(x)\varphi_y(y)\varphi_z(z),
\end{align}
where
\begin{align}
	\varphi_z(z)=\frac{1}{(\pi l_z^2)^{1/4}}e^{-z^2/(2l_z^2)},\ \varphi_y(y)=\frac{1}{(\pi l_y^2)^{1/4}}e^{-y^2/(2l_y^2)}
\end{align}
are the ground states along the $z$- and $y$-direction, respectively. $l_z$, $l_y$ denote the oscillator lengths. In the following, we will write $\textbf{r}$ to denote $\textbf{r}_\perp$ in two dimensions and $x$ in one dimension, respectively.

Now consider the incoherent region. For terms in $\hat{\phi}$ that contain an odd single particle excited state the integration in (\ref{3rdterm}) will vanish and the contribution of these terms is hence zero. Moreover, those states that are along one of the reduced axes of at least second order in the single particle state have an energy larger than twice the cut-off. As two particles in the coherent region together have allways less energy than twice the  cut-off, there can not be energy conserving scattering processes including those states. Hence, we can assume that only those states in $\hat{\phi}$ that are in the single-particle ground state along the reduced axes play a role in the evaporative process.

We deduce that the $D$-dimensional master equation can be written as
\begin{widetext}
\begin{align}
	\begin{split}
		\partial_t\rho_C|_\sigma&\simeq\frac{g_D^2}{\hbar^2}\int d^D\textbf{r}\int d^D\textbf{r}'\bigg\{\left[R^{(-)}_D\left(\frac{\textbf{r}+\textbf{r}'}{2},\textbf{r}-\textbf{r}',\hat{L}_C\right)\hat{\Phi}(\textbf{r}')\rho_C,\hat{\Phi}^\dagger(\textbf{r})\right]+\left[\hat{\Phi}^\dagger(\textbf{r}'),\rho_CR^{(+)}_D\left(\frac{\textbf{r}+\textbf{r}'}{2},\textbf{r}-\textbf{r}',\hat{L}_C\right)\hat{\Phi}(\textbf{r})\right]\\
		&+\left[R^{(+)}_D\left(\frac{\textbf{r}+\textbf{r}'}{2},\textbf{r}-\textbf{r}',-\hat{L}_C\right)\hat{\Phi}^\dagger(\textbf{r}')\rho_C,\hat{\Phi}(\textbf{r})\right]+\left[\hat{\Phi}(\textbf{r}'),\rho_CR^{(-)}_D\left(\frac{\textbf{r}+\textbf{r}'}{2},\textbf{r}-\textbf{r}',-\hat{L}_C\right)\hat{\Phi}^\dagger(\textbf{r})\right]	\bigg\}.
	\end{split}
\end{align}
\end{widetext}
Here, we have the $D$-dimensional interaction strength
\begin{align}
	g_D=\begin{cases}
		4\pi\hbar^2a_\text{s}/m,&D=3\\
		\sqrt{8\pi}\hbar^2a_\text{s}/(ml_z),&D=2\\
		2\hbar^2a_\text{s}/(ml_zl_y),&D=1
		\end{cases}
\end{align}
and
\begin{align}
	\begin{split}
		R^{(-)}_D\left(\textbf{u},\textbf{v},\hat{L}_C\right)&=\pi\int_I \frac{d^D\textbf{K}}{(2\pi)^D}\left[1+F_D\left(\textbf{u},\textbf{K}\right)\right]e^{i\textbf{K}\cdot\textbf{v}}\\
		&\times\delta(\omega(\textbf{u},\textbf{K})-\hat{L}_C/\hbar),\\
		R^{(+)}_D\left(\textbf{u},\textbf{v},\hat{L}_C\right)&=\pi\int_I \frac{d^D\textbf{K}}{(2\pi)^D}F_D\left(\textbf{u},\textbf{K}\right)e^{i\textbf{K}\cdot\textbf{v}}\\
		&\times\delta(\omega(\textbf{u},\textbf{K})-\hat{L}_C/\hbar)
	\end{split}
\end{align}
with the Wigner function describing the incoherent atoms in the ground state along the reduced axes
\begin{align}
	\begin{split}
	&F_D(\textbf{u},\textbf{K})\\
	&=\frac{\Theta(K-k_\text{cut})}{\exp\left(\beta\left[2n_\text{tot}(\textbf{u})g_D+\frac{\hbar^2\textbf{K}^2}{2m}+V_D(\textbf{u})-\mu_D\right]\right)-1}.
	\end{split}
\end{align}
Here, $V_D(\textbf{u})$ denotes the potential along the unreduced axes.

We can argue as before, approximating
\begin{align}
	\begin{split}
		&R_D^{(+)}(\textbf{u},\textbf{v},\hbar\omega)\simeq R_D^{(+)}(\textbf{u},\textbf{v},\hbar\omega_\text{cut}),\\ &R_D^{(-)}(\textbf{u},\textbf{v},\hbar\omega)\simeq R_D^{(+)}(\textbf{u},\textbf{v},\hbar\omega_\text{cut})[1+\beta\hbar\{\omega-\mu/\hbar\}]
	\end{split}
\end{align}
to derive
\begin{widetext}
\begin{align}
	\label{nD}
	\begin{split}
		i\hbar (\textbf{S})d\psi(\textbf{r})|_{\sigma}&=\mathcal{P}\bigg\{-i\int d^D\textbf{r}'\text{Re}\{\sigma_D(\textbf{r},\textbf{r}')\}\psi(\textbf{r})\psi^*(\textbf{r}')[L_D-\mu_D]\psi(\textbf{r}')+3\int d^D\textbf{r}'\text{Im}\{\sigma_D(\textbf{r},\textbf{r}')\}\psi(\textbf{r})\psi^*(\textbf{r}')[L_D-\mu_D]\psi(\textbf{r}')\\
		&-4\hbar\psi(\textbf{r})\int d^D\textbf{r}'\text{Re}\{\sigma_D(\textbf{r},\textbf{r}')\}\nabla'\cdot\textbf{j}(\textbf{r}')\bigg\}dt+\mathcal{P}\left\{i\hbar\psi dw_D+\hbar\psi dU_D\right\}.
	\end{split}
\end{align}
\end{widetext}
The noises fulfill the correlations
\begin{align}
	\begin{split}
		&\langle dw_D(\textbf{r})dw_D^*(\textbf{r}')\rangle=\frac{2k_\text{B}T}{\hbar}\sigma_D(\textbf{r},\textbf{r}')dt,\\
		&\langle dU_D(\textbf{r})dU_D(\textbf{r}')\rangle=\frac{4k_\text{B}T}{\hbar}\sigma_D(\textbf{r},\textbf{r}')dt,\\
		&\langle dU_D(\textbf{r})dw_D^*(\textbf{r}')\rangle=-\frac{2k_\text{B}T}{\hbar}\sigma_D(\textbf{r},\textbf{r}')dt.
	\end{split}
\end{align}
Here,
\begin{align}
	\begin{split}
	\sigma_D(\textbf{r},\textbf{r}')&=a_\text{s}^2\lambda_\text{th}^2N_\text{cut}\psi^*(\textbf{r})\psi(\textbf{r}')\\
	&\times\begin{cases}4k_\text{cut}(\textbf{u})\text{sinc}\left(k_\text{cut}\left(\textbf{u}\right)|\textbf{r}-\textbf{r}'|\right),&D=3\\
		2J_0(k_\text{cut}(\textbf{u})v)/l_z^2,&D=2\\
		\frac{2}{\pi}\cos(k_\text{cut}(u)v)/[k_\text{cut}(u)l_z^2l_y^2],&D=1
\end{cases}
\end{split}
\end{align}
and
\begin{align}
	\begin{split}
	L_D\psi&=\left[2n_{I,D}(\textbf{r})g_D+g_D|\psi|^2+V_D(\textbf{r})-\frac{\hbar^2}{2m}\Delta\right]\psi,\\
	n_{I,2}(\textbf{r})&=\int dzn_I(\textbf{r})\varphi_z(z)^2,\\
	n_{I,1}(\textbf{r})&=\int dydzn_I(\textbf{r})\varphi_y(y)^2\varphi_z(z)^2.
	\end{split}
\end{align}
Moreover, we have the chemical potential
\begin{align}
	\mu_D=\begin{cases}
		\mu,&D=3\\
		\mu-\hbar\omega_z/2,&D=2\\
		\mu-\hbar\omega_z/2-\hbar\omega_y/2,&D=1
	\end{cases},
\end{align}
where $\omega_j$ denotes the trapping frequency along the $j$-direction
\begin{align}
	\omega_j=\frac{\hbar}{ml_j^2}.
\end{align}

We derived a dimensionally reduced version of the evaporative damping. {In contrast to the dimensional reduction performed for energy and number damping}\cite{bradley_low-dimensional_2015}{ this reduction stays valid in traps deeper than the thermal energy $k_\text{B}T$. As long as the thermal cloud stays three dimensional, number and energy damping include interactions with all the particles in the incoherent region}\cite{bradley_low-dimensional_2015}{, while evaporative damping only includes scattering with particles in the ground state along the reduced dimensions. However, the trapping implies an increase of the coherent region density relative to the incoherent region density. As more of the physics of evaporative damping happens in the coherent region compared to number and energy damping, in reduced dimensions, evaporative damping can still be expected to play a relevant role. For deep traps in which almost all particles sit in the ground state along the reduced dimension, it might actually be enhanced. In appendix }\ref{1D}{ we estimate the strength of evaporative damping in one dimension at high phase space density and compare it with energy damping, indicating indeed a relevant role of evaporative damping in one dimension.}

\section{Conclusions}
We derive an evaporative damping mechanism acting on the classical wave in a BEC. It describes scattering processes between particles in the c-field, in which one of the particles gains enough energy to leave the coherent region. The main result, equation (\ref{dpsidrift}), completes the SPGPE finite temperature theory of ultracold Bose gases\cite{gardiner_stochastic_2002,gardiner_stochastic_2003,blakie_dynamics_2008}. Together with number and energy damping, all possible (two body) scattering events between particles in the coherent and incoherent region are being considered (see figure \ref{fig:overview}). While previously claimed to be only a minor correction in the thermalization and hence neglected, we demonstrate that in three dimensional gases evaporative damping is comparable in magnitude to the most widely applied number damping.

 The process can be decomposed into two contributions: one is akin to number damping, describing particles that are scattered out of the coherent region by another coherent region particle (as opposed to a thermal particle in number damping). The other one is akin to energy damping, describing the cooling of a particle by scattering another particle out of the coherent region. This interpretation becomes more apparent in a simplified pseudo-local form (\ref{pseudo-local}). Under further approximations, the number-damping-like contribution takes excactly the same form as the usual number damping process, only differing by the pre-factor. This gives some justification to the common practice of artificially increasing the number damping strength to match experiments \cite{weiler_spontaneous_2008,kobayashi_thermal_2016,liu_dynamical_2018,liu_kibble-zurek_2020,thudiyangal_universal_2024,comaron_quench_2019,roy_finite-temperature_2021,roy_finite-temperature_2023}. We stress, however, that at high-phase space density energy damping dominates over both number damping \cite{mehdi_mutual_2023,krause_thermal_2024,krause_equilibrium_2025} and evaporative damping (see appendix \ref{Linearization}).
 
 {A rough estimate of the evaporative damping strength close to the transition point }(\ref{sigmanl1}){ suggests that it is particularly strong compared to number damping for weakly interacting systems, for which the scattering length is much smaller than the de-Broglie wavelength. To verify the importance of evaporative damping in this regime will require future research.}
 
 {In analyzing the importance of the evaporative damping process we focused on homogenous systems. We note that in traps the bulk of the coherent region and incoherent region become spatially seperated. The physics of the evaporative damping process is more strongly reliant on the coherent density and less affected by the density of the incoherent region compared to number and energy damping. Therefore, in trapped systems we can expect an increase in the relative strength of evaporative damping. Considering evaporative damping in a trap might hence be an important future direction of study.}
 
 We also derive a dimensionally reduced form of evaporative damping, equation (\ref{nD}). The reduction is valid if a deep harmonic trapping potential is applied along one (two) of the axes. It allows the study of the effect of evaporative damping on physics in reduced dimensions. One application could be the dynamics of stable non-linear excitations of the GPE, e.g. dark solitons \cite{burger_dark_1999,busch_motion_2000,becker_oscillations_2008,jackson_dark-soliton_2007} in one dimension and vortices \cite{mehdi_mutual_2023,madison_vortex_2000,abo-shaeer_observation_2001,kwon_sound_2021} and Jones-Roberts solitons \cite{jones_motions_1982,krause_thermal_2024,meyer_observation_2017,baker-rasooli_observation_2025} in two dimensions. {We note that in dimensionally reduced systems the coherent density is increased relative to the thermal cloud. Accordingly, evaporative damping could dominate thermalization in dimensionally reduced systems, in consistency with an estimate for one dimensional systems at high phase space density in appendix }\ref{1D}{.}
 
 While in this work we aim on deriving the evaporative damping mechanism acting on a classical wave, the master equation (\ref{master}) can also describe evaporative cooling of the whole Bose gas. In order to describe all particles in the trap, the cut-off should be chosen high enough so that the coherent region is made up of all the bound states. As soon as a particle leaves the trap, it essentially leaves the system, so that the bath should be assumed to be empty ($F=0$). Note that under these assumptions the evaporative damping term is the only non Hamiltonian contribution remaining in the evolution. It might be an interesting question to work out if this leads to a useful first principal description of evaporative cooling that includes the presence of coherence between particles in different states.
 
 \section*{Acknowledgement}
 We are grateful to the Dodd-Walls Centre for Photonic and Quantum Technologies for financial support.
	
	\appendix
	\section{Diffusion Mapping}
	\label{mapping}
	To map the diffusion equation (\ref{dWdiff}) to a stochastic equation for $\psi$, we first  note that the integrand can be expanded to
	\begin{widetext}
	\begin{align}
		\begin{split}
		&\sum_{n,m}\bigg[\partial_{\alpha_n}\chi_n^*(\textbf{r})\psi(\textbf{r})\psi(\textbf{r})\partial_{\alpha_m^*}\chi_m(\textbf{r}')\psi^*(\textbf{r}')\psi^*(\textbf{r}')-2\partial_{\alpha_n}\chi_n^*(\textbf{r})\psi(\textbf{r})\psi(\textbf{r})\partial_{\alpha_m}\chi_m^*(\textbf{r}')\psi^*(\textbf{r}')\psi(\textbf{r}')\\
		&-2\partial_{\alpha_n}\chi_n^*(\textbf{r})\psi^*(\textbf{r})\psi(\textbf{r})\partial_{\alpha_m}\chi_m^*(\textbf{r}')\psi(\textbf{r}')\psi(\textbf{r}')+4\partial_{\alpha_n}\chi_n^*(\textbf{r})\psi^*(\textbf{r})\psi(\textbf{r})\partial_{\alpha_m^*}\chi_m(\textbf{r}')\psi^*(\textbf{r}')\psi(\textbf{r}')\bigg]+h.c.
		\end{split}
	\end{align}
	In the following, we list the appearing derivatives with regard to Re $\alpha$, Im $\alpha$. First, Terms with only Re $\alpha$ derivatives
	\begin{align}
		\begin{split}
			&\frac{1}{2}\sum_{n,m}\frac{\partial}{\partial \text{Re}\ \alpha_n}\bigg[\text{Re}\{\chi_n^*(\textbf{r})\psi(\textbf{r})\psi(\textbf{r})\}-2\text{Re}\{\chi_n^*(\textbf{r})\psi^*(\textbf{r})\psi(\textbf{r})\}\bigg]\frac{\partial}{\partial \text{Re}\ \alpha_m}\bigg[\text{Re}\{\chi_m^*(\textbf{r}')\psi(\textbf{r}')\psi(\textbf{r}')\}-2\text{Re}\{\chi_m^*(\textbf{r}')\psi^*(\textbf{r}')\psi(\textbf{r}')\}\bigg]\\
			&+\frac{\partial}{\partial \text{Re}\ \alpha_n}\bigg[\text{Im}\{\chi_n^*(\textbf{r})\psi(\textbf{r})\psi(\textbf{r})\}+2\text{Im}\{\chi_n^*(\textbf{r})\psi^*(\textbf{r})\psi(\textbf{r})\}\bigg]\frac{\partial}{\partial \text{Re}\ \alpha_m}\bigg[\text{Im}\{\chi_m^*(\textbf{r}')\psi(\textbf{r}')\psi(\textbf{r}')\}+2\text{Im}\{\chi_m^*(\textbf{r}')\psi^*(\textbf{r}')\psi(\textbf{r}')\}\bigg]
		\end{split}
	\end{align}
	Terms with only Im $\alpha$ derivatives
	\begin{align}
		\begin{split}
			&\frac{1}{2}\sum_{n,m}\frac{\partial}{\partial \text{Im}\ \alpha_n}\bigg[\text{Re}\{\chi_n^*(\textbf{r})\psi(\textbf{r})\psi(\textbf{r})\}+2\text{Re}\{\chi_n^*(\textbf{r})\psi^*(\textbf{r})\psi(\textbf{r})\}\bigg]\frac{\partial}{\partial \text{Im}\ \alpha_m}\bigg[\text{Re}\{\chi_m^*(\textbf{r}')\psi(\textbf{r}')\psi(\textbf{r}')\}+2\text{Re}\{\chi_m^*(\textbf{r}')\psi^*(\textbf{r}')\psi(\textbf{r}')\}\bigg]\\
			&+\frac{\partial}{\partial \text{Im}\ \alpha_n}\bigg[\text{Im}\{\chi_n^*(\textbf{r})\psi(\textbf{r})\psi(\textbf{r})\}-2\text{Im}\{\chi_n^*(\textbf{r})\psi^*(\textbf{r})\psi(\textbf{r})\}\bigg]\frac{\partial}{\partial \text{Im}\ \alpha_m}\bigg[\text{Im}\{\chi_m^*(\textbf{r}')\psi(\textbf{r}')\psi(\textbf{r}')\}-2\text{Im}\{\chi_m^*(\textbf{r}')\psi^*(\textbf{r}')\psi(\textbf{r}')\}\bigg]
		\end{split}
	\end{align}
	Terms with Re $\alpha$ derivative first
	\begin{align}
		\begin{split}
			&\frac{1}{2}\sum_{n,m}\frac{\partial}{\partial \text{Re}\ \alpha_n}\bigg[\text{Re}\{\chi_n^*(\textbf{r})\psi(\textbf{r})\psi(\textbf{r})\}-2\text{Re}\{\chi_n^*(\textbf{r})\psi^*(\textbf{r})\psi(\textbf{r})\}\bigg]\frac{\partial}{\partial \text{Im}\ \alpha_m}\bigg[\text{Im}\{\chi_m^*(\textbf{r}')\psi(\textbf{r}')\psi(\textbf{r}')\}-2\text{Im}\{\chi_m^*(\textbf{r}')\psi^*(\textbf{r}')\psi(\textbf{r}')\}\bigg]\\
			&-\frac{\partial}{\partial \text{Re}\ \alpha_n}\bigg[\text{Im}\{\chi_n^*(\textbf{r})\psi(\textbf{r})\psi(\textbf{r})\}+2\text{Im}\{\chi_n^*(\textbf{r})\psi^*(\textbf{r})\psi(\textbf{r})\}\bigg]\frac{\partial}{\partial \text{Im}\ \alpha_m}\bigg[\text{Re}\{\chi_m^*(\textbf{r}')\psi(\textbf{r}')\psi(\textbf{r}')\}+2\text{Re}\{\chi_m^*(\textbf{r}')\psi^*(\textbf{r}')\psi(\textbf{r}')\}\bigg]
		\end{split}
	\end{align}
	Terms with Im $\alpha$ derivative first
	\begin{align}
		\begin{split}
			&\frac{1}{2}\sum_{n,m}-\frac{\partial}{\partial \text{Im}\ \alpha_n}\bigg[\text{Re}\{\chi_n^*(\textbf{r})\psi(\textbf{r})\psi(\textbf{r})\}+2\text{Re}\{\chi_n^*(\textbf{r})\psi^*(\textbf{r})\psi(\textbf{r})\}\bigg]\frac{\partial}{\partial \text{Re}\ \alpha_m}\bigg[\text{Im}\{\chi_m^*(\textbf{r}')\psi(\textbf{r}')\psi(\textbf{r}')\}+2\text{Im}\{\chi_m^*(\textbf{r}')\psi^*(\textbf{r}')\psi(\textbf{r}')\}\bigg]\\
			&+\frac{\partial}{\partial \text{Im}\ \alpha_n}\bigg[\text{Im}\{\chi_n^*(\textbf{r})\psi(\textbf{r})\psi(\textbf{r})\}-2\text{Im}\{\chi_n^*(\textbf{r})\psi^*(\textbf{r})\psi(\textbf{r})\}\bigg]\frac{\partial}{\partial \text{Re}\ \alpha_m}\bigg[\text{Re}\{\chi_m^*(\textbf{r}')\psi(\textbf{r}')\psi(\textbf{r}')\}-2\text{Re}\{\chi_m^*(\textbf{r}')\psi^*(\textbf{r}')\psi(\textbf{r}')\}\bigg]
		\end{split}
	\end{align}
	
	From these expressions we see that the stochastic equation for Re $\alpha_n$, Im $\alpha_n$ are given by
	\begin{align}
		\begin{split}
		(\textbf{S})d\text{Re}\ \alpha_n|_{\sigma,\text{diff}}&=\int d^3\textbf{r}'[\text{Re}\{\chi_n^*(\textbf{r}')\psi(\textbf{r}')\psi(\textbf{r}')\}-2\text{Re}\{\chi_n^*(\textbf{r}')\psi^*(\textbf{r}')\psi(\textbf{r}')\}]dW_1(\textbf{r}')\\
		&+\int d^3\textbf{r}'[\text{Im}\{\chi_n^*(\textbf{r}')\psi(\textbf{r}')\psi(\textbf{r}')\}+2\text{Im}\{\chi_n^*(\textbf{r}')\psi^*(\textbf{r}')\psi(\textbf{r}')\}]dW_2(\textbf{r}'),\\
		(\textbf{S})d\text{Im}\ \alpha_n|_{\sigma,\text{diff}}&=\int d^3\textbf{r}'[\text{Im}\{\chi_n^*(\textbf{r}')\psi(\textbf{r}')\psi(\textbf{r}')\}-2\text{Im}\{\chi_n^*(\textbf{r}')\psi^*(\textbf{r}')\psi(\textbf{r}')\}]dW_1(\textbf{r}')\\
		&+\int d^3\textbf{r}'[\text{Re}\{\chi_n^*(\textbf{r}')\psi(\textbf{r}')\psi(\textbf{r}')\}+2\text{Re}\{\chi_n^*(\textbf{r}')\psi^*(\textbf{r}')\psi(\textbf{r}')\}]dW_2(\textbf{r}').
		\end{split}
	\end{align}
\end{widetext}
	Here
	\begin{align}
		\begin{split}
		&\langle dW_i(\textbf{r})dW_j(\textbf{r}')\rangle\\
		&=\delta_{ij}\frac{1}{2\pi}\frac{mg^2}{\hbar^3}k_\text{cut}(\textbf{u})N_\text{cut}\text{sinc}\left(k_\text{cut}|\textbf{r}-\textbf{r}'|\right)dt.
		\end{split}
	\end{align}
	Multiplying the real part equation with $\chi_n(\textbf{r})$ and the imaginary part equation with $i\chi_n(\textbf{r})$, adding the equations and summing ove $n$ we find equation (\ref{dpsidiff}).
	
	\section{Energy conservation and typical energy scale}
	\label{energycon}
	To see the effect of  energy conservation in the scattering events described by (\ref{Diffusion}), (\ref{Drift}) note
	\begin{align}
		\begin{split}
		&R^{(+)}(\textbf{u},\textbf{v},\hat{L}_C)\hat{\Phi}(\textbf{r}')\\
		&=\pi\int_I \frac{d^3\textbf{K}}{(2\pi)^3}F\left(\textbf{u},\textbf{K}\right)e^{i\textbf{K}\cdot\textbf{v}}\\
		&\times\int_{-\infty}^0dte^{i(\omega(\textbf{u},\textbf{K})-\hat{L}_C/\hbar)t}\hat{\psi}^\dagger(\textbf{r}')\hat{\psi}(\textbf{r}')\hat{\psi}(\textbf{r}').
		\end{split}
	\end{align}
	The time integrand can be evaluated to
	\begin{align}
		\begin{split}
			&e^{i(\omega(\textbf{u},\textbf{K})-\hat{L}_C/\hbar)t}\hat{\psi}^\dagger(\textbf{r}')\hat{\psi}(\textbf{r}')\hat{\psi}(\textbf{r}')\\
			&=e^{i\omega(\textbf{u},\textbf{K})t}e^{iH_Ct/\hbar}\hat{\psi}^\dagger(\textbf{r}') e^{-iH_Ct/\hbar}\\
			&\times e^{iH_Ct/\hbar}\hat{\psi}(\textbf{r}')e^{-iH_Ct/\hbar}e^{iH_Ct/\hbar}\hat{\psi}(\textbf{r}')e^{-iH_Ct/\hbar}.
		\end{split}
	\end{align}
	
	\begin{figure}
		\centering
		\includegraphics[width=.7\linewidth]{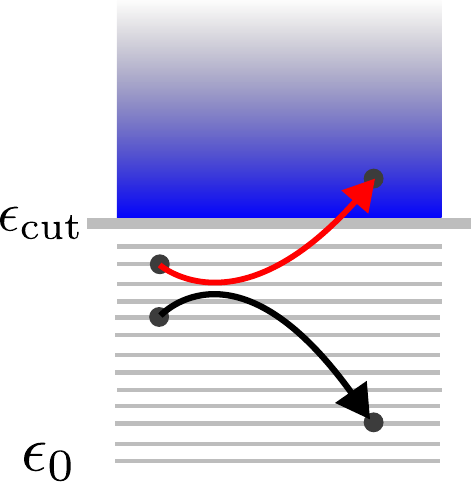}
		\caption{Schematic of the evaporative damping process. The lines symoblise the single particle states with (single particle) energies ranging from the single particle ground state energy $\epsilon_0$ to the cut-off $\epsilon_\text{cut}$. Energy conservation considerations require that, under neglect of the interaction energy, the (single particle) energies of two particles (grey dots) in the coherent region combined has to be at least $\epsilon_\text{cut}$ for an evaportive scattering to occur.}
		\label{fig:g3}
	\end{figure}
	
	If we were to approximate the time evolution by its value according to the evolution of a condensate, the integral vanishes (neglecting the principal value). This might on first sight lead to the assumption that evaporative damping would not be an important contribution to the dynamics. However, this approximation ignores that we can not expect the condensate alone to lead to evaporation. Instead, high energetic particles in the coherent region can be expected to play a major role in evaporative damping: the combined energy of two colliding particles needs to be larger than the cut-off to begin with in order to bring one of them into the incoherent region (see also figure \ref{fig:g3}). At least some of the participating particles will hence have a high energy and thus see most of the coherent region particles as essentially uncorrelated. Based on this reasoning, we approximate the interaction Hamiltonian for the coherent region in $H_C$ by the forward scattering part
	\begin{align}
		\frac{g}{2}\hat{\psi}^\dagger(\textbf{r})\hat{\psi}^\dagger(\textbf{r})\hat{\psi}(\textbf{r})\hat{\psi}(\textbf{r})\rightarrow2gn_C(\textbf{r})\hat{\psi}^\dagger(\textbf{r})\hat{\psi}(\textbf{r})
	\end{align}
	resulting in
	\begin{align}
		\begin{split}
			H_C&\simeq\int d^3\textbf{r}\left[\hat{\psi}^\dagger(\textbf{r})H_\text{sp}\hat{\psi}(\textbf{r})+2gn_\text{tot}(\textbf{r})\hat{\psi}^\dagger(\textbf{r})\hat{\psi}(\textbf{r})\right].
		\end{split}
	\end{align}
	While this approximation can be expected to be valid above the critical temperature, strictly sticking to it becomes increasingly questionable below the critical temperature. However, it can be expected to still be valid at high energetic single particle states and should suffice for deriving a typical energy scale. We can then write (for sufficiently short times and neglecting the curvature of the total density)
	\begin{align}
		\begin{split}
		&e^{iH_Ct/\hbar }\hat{\psi}^\dagger(\textbf{r}') e^{-iH_Ct/\hbar}\\
		&=\sum_j\frac{1}{j!}(it)^j[H_\text{sp}+2gn_\text{tot}(\textbf{r}')]^j\hat{\psi}^\dagger(\textbf{r}')\\
		&\simeq\sum_j\sum_n\frac{1}{j!}(it)^j[\epsilon_n+2gn_\text{tot}(\textbf{r}')]^j\chi_n^*(\textbf{r}')\hat{a}_n^\dagger\\
		&=\sum_n e^{i[\epsilon_n+2gn_\text{tot}(\textbf{r}')]t/\hbar}\chi_n^*(\textbf{r}')\hat{a}_n^\dagger
		\end{split}
	\end{align}
	and analogously
	\begin{align}
		\begin{split}
			&e^{iH_Ct/\hbar }\hat{\psi}(\textbf{r}') e^{-iH_Ct/\hbar}\\
			&\simeq\sum_n e^{-i[\epsilon_n+2gn_\text{tot}(\textbf{r}')]t/\hbar}\chi_n(\textbf{r}')\hat{a}_n.
		\end{split}
	\end{align}
	The time integral then yields energy conservation (under neglect of the principal value and approximating $\textbf{r}'\simeq\textbf{u}$ in $n_\text{tot}$)
	\begin{align}
		\hbar\omega(\textbf{u},\textbf{K})-2gn_\text{tot}(\textbf{u})=-\epsilon_n+\epsilon_m+\epsilon_l.
	\end{align}
	As
	\begin{align}
		0\leq\epsilon_n,\epsilon_m,\epsilon_l\leq\epsilon_\text{cut},\ \epsilon_\text{cut}\leq\hbar\omega(\textbf{u},\textbf{K})-2gn_\text{tot}(\textbf{u})
	\end{align}
	we have
	\begin{align}
		\epsilon_\text{cut}\leq-\epsilon_n+\epsilon_m+\epsilon_l\leq2\epsilon_\text{cut}.
	\end{align}
	Thus, the typical energy taken by $\hat{L}_C$ in (\ref{Diffusion}), (\ref{Drift}) is $\sim\epsilon_\text{cut}+2n_\text{tot}g$ and clearly not $\sim\mu$ as for numbr damping.
	
	\section{Justification of the constant energy approximation}
	\label{omegacutapp}
\begin{figure}
		\centering
		\includegraphics[width=0.7\linewidth]{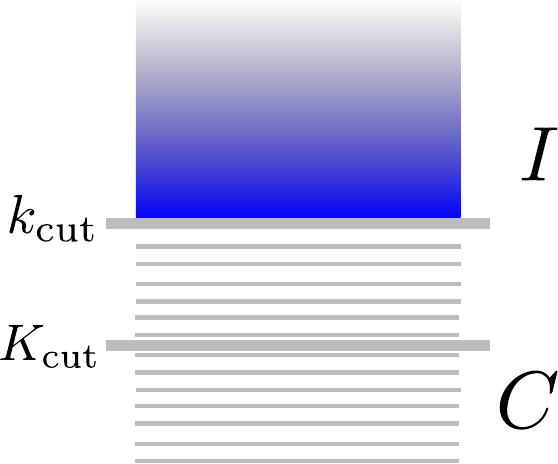}
		\caption{To estimate the effect of unphysical scattering events due to the constant energy approximation we introdue a second cut-off with momentum at the cut-off $K_\text{cut}$ for the projector. The coherent region then consists of particles that occupy states below this new cut-off, while the incoherent region encompasses still states above the original cut-off. States between these cut-offs are ignored.}
		\label{fig:kcut}
	\end{figure}

	We already saw in appendix \ref{energycon} that the typical energy scale in the scattering processes associated with evaporative damping is $\epsilon_\text{cut}$. In the following, we give some more justification for the constant energy approximation employed in the main. In particular, the approximation (\ref{Rp}) includes an implicit
	\begin{align}
		\Theta(\omega-\omega_\text{cut})\simeq1.
	\end{align}
	Although this approximation looks rather crude on first sight, the implicit momentum conservation ensures that most scattering events in which the energy loss/gain of the coherent region is less than $\hbar\omega_\text{cut}$ are still discarded after the replacement. In a homogenous gas at high phase space density we can estimate the error made in such an approximation as follows:
	\begin{itemize}
	\item An unphysical scattering event requires that some of the particles in the coherent region prior or post scattering have an energy of less then half the cut-off (neglecting the interaction energy; more precisely their energy is less than $\epsilon_\text{cut}/2+2n_\text{tot}g$).
	\item We can estimate the effect of all those scattering events of which all particles participating in the coherent region have less than half this energy by simply placing the cut-off in the projector on half the proper cut-off energy $\epsilon_\text{cut}/2$.
	\item We argue slightly more general and set the projector cut-off to an energy resulting in the momentum $K_\text{cut}$.
	\item Throughout, we keep the cut-off for the incoherent region at $k_\text{cut}$, so that some states are assigned to neither the coherent nor the incoherent region (see figure \ref{fig:kcut}).
	\end{itemize}

	According to (\ref{sigmabar}) and neglecting density fluctuations we can then write for the strength of the damping (using (\ref{thetavar}))
	\begin{align}
		\label{sigmaKcut}
		\begin{split}
		&\bar{\sigma}_\text{Kcut}\\
		&=\frac{N_\text{cut}}{\pi}\frac{\lambda_\text{th}^2}{\xi^2}\int_0^\infty \frac{dv}{\xi}\frac{v}{\xi}\sin(k_\text{cut}v)\\
		&\times\exp\left(\int \frac{d^3\textbf{k}'}{(2\pi)^3}\frac{2\pi}{n_C\lambda_\text{th}^2\xi}\frac{e^{i\textbf{k}'\cdot(\textbf{r}-\textbf{r}')}-1}{(k')^2}\Theta(K_\text{cut}-k')\right)\\
		&\simeq\frac{N_\text{cut}}{2\pi}\frac{\lambda_\text{th}^2}{\xi^2}\int_0^\infty \frac{dv}{\xi}\frac{v}{\xi}\sin(k_\text{cut}v)\left(\frac{1}{\pi n_C\lambda_\text{th}^2\xi}\frac{\text{Si}(K_\text{cut}v)}{v}\right)^2\\
		&=\frac{N_\text{cut}}{2\pi^3}\frac{1}{n_C\lambda_\text{th}^2\xi}\frac{1}{n_C\xi^3}\int_0^\infty dV\text{sinc}(V)\text{Si}\left(\frac{K_\text{cut}}{k_\text{cut}}V\right)^2.
		\end{split}
	\end{align}
	
	For $K_\text{cut}<k_\text{cut}/\sqrt{2}\equiv k_2$ the energy of the two particles in the coherent region together is less than the cut-off energy, so that all scattering events contributing to evaporative damping are unphysical (violating energy conservation). This allows us to estimate the influence of those unphysical scattering events on the damping strength:
	\begin{align}
		\frac{\bar{\sigma}_{k_2}}{\bar{\sigma}_\text{kcut}}\simeq0.1689...
	\end{align}
	In other words, the unphysical scattering events contribute with less than $17\%$ to the strength of evaporative damping.
	
	\begin{figure}
		\centering
		\includegraphics[width=1\linewidth]{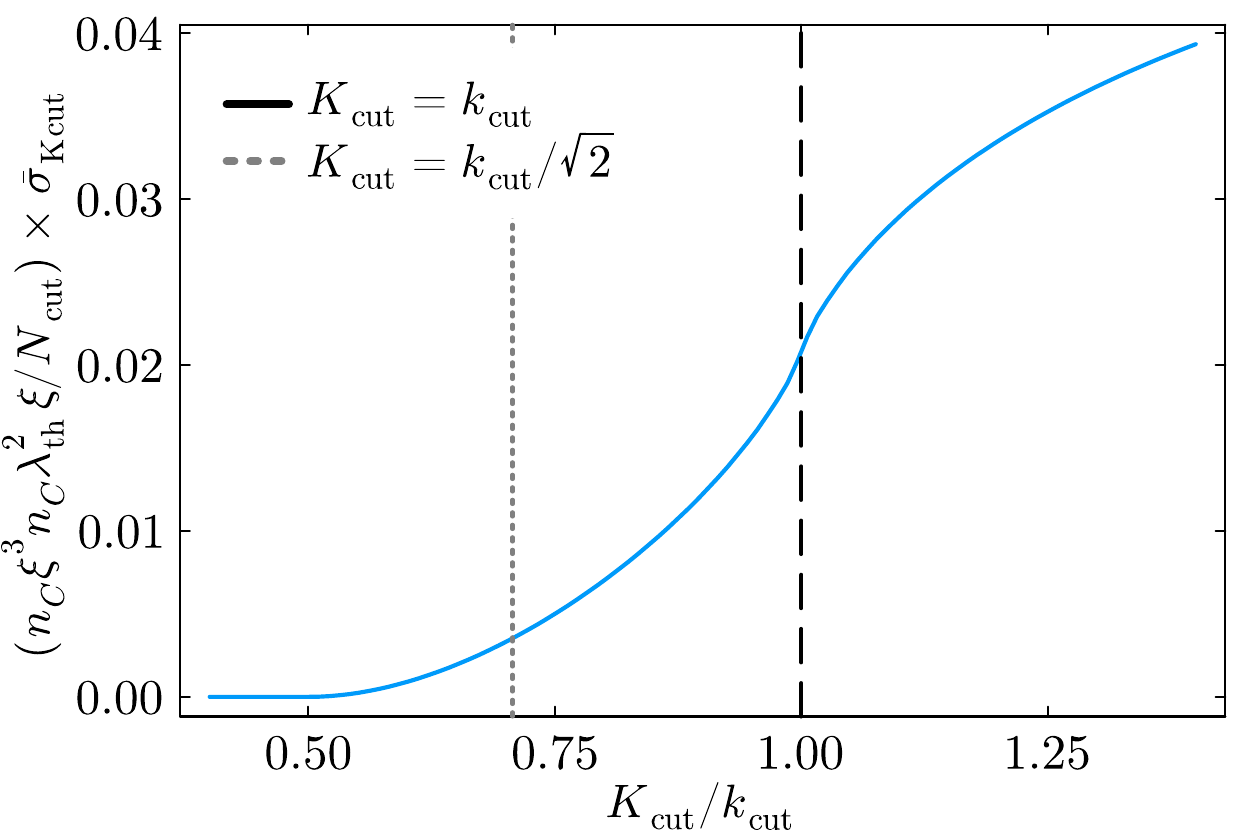}
		\caption{Estimate of the strength of evaporative damping $\bar{\sigma}_{K\text{cut}}$ for different choices of the cut-off for the coherent region $K_\text{cut}$ relative to the cut-off for the incoherent region $k_\text{cut}$. In the SPGPE the cut-offs coincide $K_\text{cut}=k_\text{cut}$, so that each state is either covered by the coherent or the incoherent region. The grey dotted line marks $K_\text{cut}=k_\text{cut}/\sqrt{2}$, so that left of it only unphysical scattering events (violating energy conservation) contribute to $\bar{\sigma}_\text{Kcut}$. For $K_\text{cut}<k_\text{cut}/2$ evaporative damping vanishes, in consistency with momentum conservation of the scattering processes. For $K_\text{cut}>k_\text{cut}$ further physical scattering events contribute to the strength of evaporative damping. Those, however, are in SPGPE treatment already accounted for by number damping.}
		\label{fig:sigmabar}
	\end{figure}
	Figure \ref{fig:sigmabar} shows $\bar{\sigma}_\text{Kcut}$ for different $K_\text{cut}$. It vanishes for $K_\text{cut}<k_\text{cut}/2$, in consistency with momentum conservation and increases monotonically with increasing $K_\text{cut}$, as more and more scattering events are included in the evaporative damping. For $K_\text{cut}>k_\text{cut}$ scattering events that are already accounted for in number damping increase the strength of evaporative damping further. As in the SPGPE $K_\text{cut}=k_\text{cut}$ such double counting does not occur.
	
	\section{Coherent number growth}
	\label{numbergrowth}
	Similiar to number damping and unlike energy damping, evaporative damping allows the exchange of particles with the thermal cloud. In this section, we estimate the growth of the coherent region induced by it. Multiplying equation (\ref{dpsidrift}) by $\psi^*$, taking the real part and integrating gives (where for simplicity we ignore the projector and noise in this section)
	\begin{align}
		\begin{split}
			\frac{dN_C}{dt}\bigg|_\sigma&=-\frac{2}{\hbar}\int d^3\textbf{r}\int d^3\textbf{r}'\text{Re}\{\sigma(\textbf{r},\textbf{r}')\}|\psi(\textbf{r})|^2\\
			&\times\text{Re}\left\{\psi^*(\textbf{r}')[L_C-\mu]\psi(\textbf{r}')\right\}\\
			&-3\int d^3\textbf{r}\int d^3\textbf{r}'\text{Im}\{\sigma(\textbf{r},\textbf{r}')\}|\psi(\textbf{r})|^2\nabla'\cdot\textbf{j}(\textbf{r}').
		\end{split}
	\end{align}
	Note that the contribution from the term we identified as energy damping like vanishes. The second term is an integral over a usually strongly fluctuating intergrand, so that we can expect that the change in particle number is essentially determined by the number damping like contribution.
	
	We can also estimate the change in grand-canonical energy
	\begin{align}
		K_C=H_C-\mu N_C.
	\end{align}
	Under the neglect of projector and noise we find
	\begin{align}
		\begin{split}
			&\frac{dK_C}{dt}\bigg|_\sigma\\
			&=-\frac{2}{\hbar}\int d^3\textbf{r}\int d^3\textbf{r}'\text{Re}\{\sigma(\textbf{r},\textbf{r}')\}\\
			&\times\text{Re}\left\{\left[\psi(\textbf{r})[L_C-\mu]\psi^*(\textbf{r})\right]\psi^*(\textbf{r}')[L_C-\mu]\psi(\textbf{r}')\right\}\\
			&+\frac{6}{\hbar}\int d^3\textbf{r}\int d^3\textbf{r}'\text{Im}\{\sigma(\textbf{r},\textbf{r}')\}\\
			&\times\text{Im}\left\{\left[\psi(\textbf{r})[L_C-\mu]\psi^*(\textbf{r})\right]\psi^*(\textbf{r}')[L_C-\mu]\psi(\textbf{r}')\right\}\\
			&-4\hbar\int d^3\textbf{r}\int d^3\textbf{r}'\text{Re}\{\sigma(\textbf{r},\textbf{r}')\}\nabla\cdot\textbf{j}(\textbf{r})\nabla'\cdot\textbf{j}(\textbf{r}').
		\end{split}
	\end{align}
	
	\section{Linearization}
	\label{Linearization}
	In order to gain a better understanding of the importance of evaporative damping, we perform a linearization along the lines of \cite{krause_equilibrium_2025}. We consider a three dimensional box at high-phase space density and linearize (\ref{dpsidrift}) via approximating
	\begin{align}
		\psi\simeq\sqrt{n_C}\left(1+\frac{\delta n}{2}\right)e^{i\theta}
	\end{align}
	and assuming
	\begin{align}
		\delta n,\nabla\theta\ll1,\ \mu\simeq n_Cg.
	\end{align}
	We derive
	\begin{align}
		\begin{split}
			&d\left(\frac{\delta n(\textbf{r})}{2}+i\theta(\textbf{r})\right)\bigg|_\sigma\simeq\mathcal{P}\bigg\{\int d^3\textbf{r}'n_C\sigma(\textbf{r}-\textbf{r}')\\
			&\times\bigg[-\delta n(\textbf{r}')+\frac{\xi^2}{4}\Delta'\delta n(\textbf{r}')+\frac{9}{2}i\xi^2\Delta'\theta(\textbf{r}')\bigg]\frac{n_Cg dt}{\hbar}\\
			&+dW-2dW^*\bigg\}.
		\end{split}
	\end{align}
	In the following we work in dimensionless units by replacing $t\rightarrow \hbar t/[n_Cg]$, $k\rightarrow k/\xi$, $r\rightarrow \xi r$. The dimensionless function $\sigma$ can be approximated by (we write $\sigma$ as a function of a single variable as in this regime it depends only on $\textbf{v}$ and neglect the density fluctuations considering that phase fluctuations dominate \cite{krause_equilibrium_2025})
	\begin{align}
		\begin{split}
			&\sigma(\textbf{r}-\textbf{r}')\simeq\frac{N_\text{cut}}{(2\pi)^2}\frac{k_\text{cut}\xi}{n_C\xi^3}\frac{\lambda_\text{th}^2}{\xi^2}\text{sinc}(k_\text{cut}|\textbf{r}-\textbf{r}'|)\\
			&\times\exp\left(\int \frac{d^3\textbf{k}'}{(2\pi)^3}\frac{2\pi}{n_C\lambda_\text{th}^2\xi}\frac{e^{i\textbf{k}'\cdot(\textbf{r}-\textbf{r}')}-1}{(k')^2}\Theta(k_\text{cut}-k')\right),\\
			&\langle dW(\textbf{r})dW^*(\textbf{r}')\rangle=\frac{4\pi}{n_C\lambda_\text{th}^2\xi}\sigma(\textbf{r}-\textbf{r}')\frac{n_Cgdt}{\hbar}.
		\end{split}
	\end{align}
	Performing a Fourier transform
	\begin{align}
		\mathcal{F}\{\circ\}(\textbf{k})=\int d^3\textbf{r}e^{-i\textbf{k}\cdot\textbf{r}}\circ
	\end{align}
	we find ($dW_1=-\mathcal{F}\{\text{Re}\{dW\}\}$, $dW_2=\mathcal{F}\{\text{Im}\{dW\}\}$)
	\begin{align}
		\label{linF}
		d\begin{bmatrix}
			\delta\tilde{n}(\textbf{k}) \\
			\tilde{\theta}(\textbf{k})
		\end{bmatrix}\bigg|_\sigma=-A(\textbf{k})\begin{bmatrix}
			\delta\tilde{n}(\textbf{k}) \\
			\tilde{\theta}(\textbf{k})
		\end{bmatrix}dt+B(\textbf{k})\begin{bmatrix}
			dW_1(\textbf{k})\\
			dW_2(\textbf{k})
		\end{bmatrix}.
	\end{align}
	Here,
	\begin{align}
		\begin{split}
			A(\textbf{k})&=\begin{bmatrix}
				2n_C\xi^3\tilde{\sigma}(\textbf{k})(1+k^2/4) & 0 \\
				0 & \frac{9}{2}n_C\xi^3\tilde{\sigma}(\textbf{k})k^2
			\end{bmatrix},\\
			&B(\textbf{k})=\begin{bmatrix}
				2(2\pi)^2\sqrt{\frac{\tilde{\sigma}(\textbf{k})}{n_C\lambda_\text{th}^2\xi}} & 0\\
				0 & 3(2\pi)^2\sqrt{\frac{\tilde{\sigma}(\textbf{k})}{n_C\lambda_\text{th}^2\xi}}
			\end{bmatrix},\\
			&\langle dW_i^*(\textbf{k},t)dW_j(\textbf{k}',t)\rangle=\delta_{ij}\delta(\textbf{k}-\textbf{k}')dt
		\end{split}
	\end{align}
	leading to the correct equilibrium distributions for $\delta\tilde{n}$, $\tilde{\theta}$.
	
	We can calculate $\tilde{\sigma}$ in this regime:
	\begin{align}
		\begin{split}
			&n_C\xi^3\tilde{\sigma}(\textbf{k})\\
			&=\frac{1}{\pi}\frac{N_\text{cut}}{k_\text{cut}\xi k}\frac{\lambda_\text{th}^2}{\xi^2}\int_0^\infty dy\\
			&\exp\left(\frac{1}{\pi}\frac{k_\text{cut}}{n_C\lambda_\text{th}^2}\frac{1}{y}[\text{Si}(y)-y]\right)\sin(ky/[k_\text{cut}\xi])\sin(y)\\
			&\simeq\frac{1}{2\pi}\frac{N_\text{cut}}{k_\text{cut}\xi k}\frac{\lambda_\text{th}^2}{\xi^2}\int_0^\infty dy\left(\frac{1}{\pi}\frac{k_\text{cut}}{n_C\lambda_\text{th}^2}\frac{\text{Si}(y)}{y}\right)^2\\
			&\times\sin(ky/[k_\text{cut}\xi])\sin(y).\\
			&=\frac{1}{2\pi^3}\frac{N_\text{cut}}{n_C\lambda_\text{th}^2\xi n_C\xi^3}\int_0^\infty dy\text{sinc}(y)\text{sinc}(ky/[k_\text{cut}\xi])\text{Si}(y)^2.
		\end{split}
	\end{align}
	Note that
	\begin{align}
		\bar{\sigma}=n_C\xi^3\tilde{\sigma}(0).
	\end{align}
	The decay rate of the occupation at momentum $k$ (see equation (41) in \cite{krause_equilibrium_2025}) hence will be
	\begin{align}
		\begin{split}
			\Gamma_k|_\sigma&=\frac{1}{48}\frac{N_\text{cut}}{n_C\lambda_\text{th}^2\xi n_C\xi^3}\left(1+\frac{5}{2}k^2\right)\frac{12}{\pi^3}\\
			&\times\int dy\text{sinc}(y)\text{sinc}(ky/k_\text{cut})\text{Si}(y)^2\\
			&\simeq\frac{1}{48}\frac{N_\text{cut}}{n_C\lambda_\text{th}^2\xi n_C\xi^3}\left(1+\frac{5}{2}k^2\right).
		\end{split}
	\end{align}
	In particular
	\begin{align}
		\lim_{k\rightarrow0}\Gamma_k|_\sigma=\frac{1}{48}\frac{N_\text{cut}}{n_C\lambda_\text{th}^2\xi n_C\xi^3}=\bar{\sigma}(\textbf{r}).
	\end{align}
	
	Compared to the decay rate induced by energy damping we have
	\begin{align}
		\frac{\Gamma_k|_\varepsilon}{\Gamma_k|_\sigma}\simeq\frac{24}{\pi}n_C\lambda_\text{th}^2\xi\frac{k}{1+5k^2/2}.
	\end{align}
	Except for low momenta, energy damping hence dominates the diffusive dynamics at high phase space density.
	
	\section{One Dimensional Reduced Dynamics at high Phase Space Density}
	\label{1D}
	In the one dimensional case we can calculate $\bar{\sigma}$ analogously to the three dimensional case outlined in \ref{Approximate}. At high phase space density in a box we find
	\begin{align}
		\begin{split}
			\bar{\sigma}_1&=n_C\frac{2}{\pi}\frac{a_\text{s}^2\lambda_\text{th}^2N_\text{cut}}{k_\text{cut}l_z^2l_y^2}\int dv\cos(k_\text{cut}v)\\
			&\times\langle\psi^*(x+v/2)\psi(x-v/2)\rangle\\
			&\simeq\frac{2}{\pi}\frac{n_Ca_\text{s}^2\lambda_\text{th}^2N_\text{cut}}{k_\text{cut}l_z^2l_y^2}\int dv\cos(k_\text{cut}v)\\
			&\times n_C\exp\left(\left\langle\theta(x+v/2)\theta(x-v/2)-\theta(x)^2\right\rangle\right)\\
			&\simeq\frac{2}{\pi}\frac{n_C^2a_\text{s}^2\lambda_\text{th}^2N_\text{cut}}{k_\text{cut}l_z^2l_y^2}\int dv\cos(k_\text{cut}v)\\
			&\times\exp\left(\frac{1}{n_C\lambda_\text{th}^2}\int_{-k_\text{cut}}^{k_\text{cut}}dk\frac{\cos(kv)-1}{k^2}\right)\\
			&=\frac{2}{\pi}\frac{n_C^2a_\text{s}^2\lambda_\text{th}^2N_\text{cut}}{k_\text{cut}^2l_z^2l_y^2}\int dy\cos(y)\\
			&\times\exp\left(\frac{2}{n_C\lambda_\text{th}^2}\frac{1-\cos(y)-\text{Si}(y)y}{k_\text{cut}}\right)
		\end{split}
	\end{align}
	For $n_C\lambda_\text{th}^2k_\text{cut}\ll1$, as is the case at high phase space density and SPGPE validity, we can approximate the argument of the exponential by its asymptotic and find
	\begin{align}
		\begin{split}
			\bar{\sigma}_1&\simeq\frac{2}{\pi}\frac{n_C^2a_\text{s}^2\lambda_\text{th}^2N_\text{cut}}{k_\text{cut}^2l_z^2l_y^2}\int dy\cos(y)\exp\left(\frac{-\pi y}{n_C\lambda_\text{th}^2k_\text{cut}}\right)\\
			&=4\frac{n_Ca_\text{s}^2N_\text{cut}}{k_\text{cut}^3l_z^2l_y^2}.
		\end{split}
	\end{align}
	For a symmetric trap $l_y=l_z$, we have $l_z^2=2n_Ca_\text{s}\xi^2$ \cite{bradley_low-dimensional_2015} and hence conclude
	\begin{align}
		\bar{\sigma}_1\simeq\frac{N_\text{cut}}{[k_\text{cut}\xi]^3n_C\xi}.
	\end{align}
	
	In one dimension, energy damping can be cast into a local form \cite{bradley_low-dimensional_2015} allowing a direct comparison to the energy damping like term in the (pseudo-)local form of evaporative damping (\ref{pseudo-local}), as the processes than only differ in the prefactor. The energy damping strength is given by
	\begin{align}
		\bar{\mathcal{M}}=2\sqrt{8\pi}\frac{a_\text{s}^2}{l_z}N_\text{cut}=4\sqrt{\pi}\frac{a_\text{s}^2N_\text{cut}}{\sqrt{n_Ca_\text{s}}\xi},
	\end{align}
	while the evaporative damping strength for the energy damping like term is determined by $4\bar{\sigma}_1/n_C$. The relative strength between energy and evaporative damping is hence given by
	\begin{align}
		\frac{n_C\bar{\mathcal{M}}}{4\bar{\sigma}_1}=\sqrt{\pi}[n_Ca_\text{s}]^{3/2}[k_\text{cut}\xi]^3.
	\end{align}
	With the ideal cut-off choice identified in \cite{krause_equilibrium_2025} we can write this as
	\begin{align}
		\frac{n_C\bar{\mathcal{M}}}{4\bar{\sigma}_1}=\sqrt{\pi}[n_Ca_\text{s}]^{3/2}\left[2\left(2\frac{k_\text{B}T}{\hbar\omega_z}\frac{1}{n_Ca_\text{s}}\right)^{1/3}-2\right]^{3/2}.
	\end{align}
	Energy damping maximizes compared to evaporative damping for
	\begin{align}
		n_Ca_\text{s}=2(2/3)^3\frac{k_\text{B}T}{\hbar\omega_z}
	\end{align}
	in which case it is
	\begin{align}
		\frac{n_C\bar{\mathcal{M}}}{4\bar{\sigma}_1}=2^{3/2}\left(\frac{2}{3}\right)^{9/2}\sqrt{\pi}\left(\frac{k_\text{B}T}{\hbar\omega_z}\right)^{3/2}\simeq0.8\left(\frac{k_\text{B}T}{\hbar\omega_z}\right)^{3/2}.
	\end{align}
	For the dimensional reduction to be valid, the thermal energy $k_\text{B}T$ should not be much larger than the transverse trap energy scale $\hbar\omega_z$. We conclude that in one dimensional systems, evaporative damping will often be at least comparable to energy damping and increases in importance with the depth of the transverse trap.
	
	In two dimensions the slow algebraic decay of first order correlations\cite{hadzibabic_two-dimensional_2011} means that the derivation of the (pseudo-)local form becomes invalid. Hence, it hampers an analytic estimate of the strength of evaporative damping along the lines of the estimates in one and three dimensions.

	\end{document}